\documentclass[prl,reprint,aps,footinbib]{revtex4-2}

\usepackage[utf8]{inputenc}
\usepackage{braket}
\usepackage{amsmath}
\usepackage{amssymb}
\usepackage{graphicx}
\usepackage[english]{babel}
\usepackage{hyperref}
\usepackage{array}
\usepackage{booktabs}
\usepackage{multirow}
\usepackage{enumitem}
\usepackage{physics}
\usepackage[loose]{units}
\usepackage{natbib}

\newcommand{\half}{\ensuremath{\frac{1}{2}}}
\newcommand{\iu}{\ensuremath{i\mkern1mu}}
\DeclareMathOperator{\atantwo}{atan2}

\newcommand{\spinproj}[2]{\ensuremath{F_{#1}^{#2}}}
\newcommand{\Fxlab}{\spinproj{x}{0}}

\newcommand{\Fxone}{\spinproj{x}{1}}
\newcommand{\Fxoner}{\spinproj{x}{\text{1R}}}

\newcommand{\Fxtwos}{\spinproj{x}{\text{2S}}}
\newcommand{\Fyone}{\spinproj{y}{1}}
\newcommand{\Fyoner}{\spinproj{y}{\text{1R}}}

\newcommand{\Fytwos}{\spinproj{y}{\text{2S}}}
\newcommand{\Fzone}{\spinproj{z}{1}}
\newcommand{\Fzoner}{\spinproj{z}{\text{1R}}}
\newcommand{\Fzoneri}{\spinproj{z}{\text{1R}}\ensuremath{(t=0)}}

\newcommand{\Fztwos}{\spinproj{z}{\text{2S}}}
\newcommand{\rabict}{\ensuremath{\Omega_\text{c}(t)}}
\newcommand{\rabic}{\ensuremath{\Omega_\text{c}}}
\newcommand{\freqc}{\ensuremath{\omega_\text{c}}}
\newcommand{\freql}{\ensuremath{\omega_\text{L}}}
\newcommand{\detc}{\ensuremath{\Delta_\text{c}}}
\newcommand{\rabis}{\ensuremath{\Omega_\text{s}}}
\newcommand{\freqs}{\ensuremath{\omega_\text{s}}}
\newcommand{\phis}{\ensuremath{\phi_\text{s}}}
\newcommand{\phic}{\ensuremath{\phi_\text{c}}}
\newcommand{\bs}{\ensuremath{B_\text{s}}}
\newcommand{\phixy}{\ensuremath{\phi_{xy}^\text{2S}}}
\newcommand{\sigz}{\ensuremath{\hat{\sigma}_\text{z}}}
\newcommand{\sigy}{\ensuremath{\hat{\sigma}_\text{y}}}
\newcommand{\sigx}{\ensuremath{\hat{\sigma}_\text{x}}}
\newcommand{\hlabt}{\ensuremath{\hat{H}_0(t)}}

\newcommand{\hone}{\ensuremath{\hat{H}_1}}
\newcommand{\hione}{\ensuremath{\hat{H}_\text{1I}}}
\newcommand{\honer}{\ensuremath{\hat{H}_\text{1R}}}

\newcommand{\htwo}{\ensuremath{\hat{H}_{2}}}
\newcommand{\htwos}{\ensuremath{\hat{H}_\text{2S}}}

\begin{document}
	\title{Quantum spectral analysis by continuous measurement of Landau-Zener transitions}
	\author{Christopher~C.~Bounds}
	\email{christopher.bounds@monash.edu}
	\author{Josh~P.~Duff}
	\author{Alex~Tritt}
	\author{Hamish~A.\,M.~Taylor}
	\author{George~X.~Coe}
	\author{Sam~J.~White}
	\author{L.\,D.~Turner}
	\affiliation{School of Physics and Astronomy, Monash University, Melbourne, Victoria 3800, Australia}
	
	\begin{abstract}
		We demonstrate the simultaneous estimation of signal frequency and amplitude by a single quantum sensor in a single experimental shot.
		Sweeping the qubit splitting linearly across a span of frequencies induces a non-adiabatic Landau-Zener transition as the qubit crosses resonance.
		The signal frequency determines the time of the transition, and the amplitude its extent.
		Continuous weak measurement of this unitary evolution informs a parameter estimator retrieving precision measurements of frequency and amplitude.
		Implemented on radiofrequency-dressed ultracold atoms read out by a Faraday spin-light interface, we sense a magnetic signal with $\unit[20]{pT}$ precision in amplitude, and near-transform-limited precision in frequency, in a single $\unit[300]{ms}$ sweep from $7$ to $\unit[13]{kHz}$.
		The protocol realizes a swept-sine quantum spectrum analyzer, potentially sensing hundreds or thousands of channels with a single quantum sensor.
	\end{abstract}
	\maketitle
	
	The quest for ever more sensitive measurements, especially of signals emanating from micro- and nanoscale sources, leads inexorably to quantum sensors free from thermal noise and calibration drift~\cite{degen_quantum_2017}.
	Instruments that measure directly in the frequency domain -- spectrum, signal or wave analyzers -- have been central to advances in physics, engineering~\cite{engelson_modern_1984} and speech science~\cite{koenig_sound_1946} amongst many other fields, for over a century~\cite{thomson_iv_1878,michelson_new_1898}.
	A quantum spectrum analyzer is a quantum sensor measuring directly in the frequency domain, indicating both the amplitude and frequency of harmonic components within a frequency span.
	To date this has been realized by either incoherent noise spectroscopy~\cite{kotler_nonlinear_2013} or coherent filter banks, where each measurement senses a distinct, fixed frequency.
	Continuous arrays of qubits formed by large ensembles in magnetic bias gradients have demonstrated multi-gigahertz spans but with intrinsic and inhomogeneous broadening limiting resolution to order $\unit[1]{MHz}$~\cite{chipaux_wide_2015,magaletti_quantum_2022}.
	Discrete arrays of qubits have the potential to achieve higher spectral resolution and sensitivity in reduced volume by using dynamical decoupling to render each qubit sensitive to a single frequency~\cite{haeberlen_coherent_1968,facchi_unification_2004,cai_robust_2012,ajoy_quantum_2017,schmitt_submillihertz_2017,boss_quantum_2017-1,trypogeorgos_synthetic_2018,anderson_continuously_2018,staudenmaier_phase-sensitive_2021,meinel_heterodyne_2021}, harmonics series~\cite{khodjasteh_dynamically_2009,ban_photon-echo_1998,taylor_high-sensitivity_2008,kotler_single-ion_2011} or frequency band~\cite{frey_application_2017}, and insensitive elsewhere.
	Analyzing a wide span at high resolution is a formidable scaling challenge for this approach, requiring many qubits with long coherence times, individually addressed for quantum control, and individually read out.
	To date such discrete filter banks have sensed four distinct frequencies~\cite{zhang_selective_2017}.
	In this Letter, we present an alternative approach to quantum spectrum analysis, with a spectral resolution equivalent to a filter bank with sixty channels, realized in a single-shot measurement of a single sensor under homogeneous control.
	
	Instead of parallelizing across many qubits each sensing at a fixed frequency, we subject a single sensor comprising many atoms to time-dependent but spatially homogeneous control.
	We continuously sweep the sensing frequency of this single long-decoherence-time sensor across the span, while making a continuous weak measurement~\cite{jasperse_continuous_2017} of its unitary evolution.
	An unknown oscillating magnetic field within the frequency span induces a Landau-Zener transition between the diabatic eigenstates.
	Continuous measurement reveals the resonance time and hence signal frequency, while the deviation from the initial state encodes interaction strength and hence signal amplitude.
	Landau-Zener transitions are widely used in quantum control to generate rotations insensitive to shifts in resonance~\cite{ivakhnenko_nonadiabatic_2023}.
	Here we show that Landau-Zener transitions open a new perspective on quantum measurement. 
	
	Our sensing qubit consists of Zeeman states dressed by resonant radiofrequency radiation.
	The dressed state splitting is simply the Rabi frequency of this drive, and we realize the Landau-Zener spectrum analyzer by linearly increasing the drive amplitude over time.
	Dressed states have two important advantages for time-dependent sensing.
	First, dressing realizes continuous dynamical decoupling~\cite{facchi_unification_2004,cai_robust_2012,trypogeorgos_synthetic_2018,anderson_continuously_2018} from magnetic noise outside the span, including static detuning error and low-frequency technical noise.
	Second, the on-going Rabi flopping orthogonal to the Larmor precession enables a single-axis weak measurement to perform continuous state tomography~\cite{silberfarb_quantum_2005} on the evolving dressed qubit.
	
	The protocol can be understood by considering a spin-$\half$ system with laboratory frame Hamiltonian $\hlabt = -\hbar\freql \sigz / 2 + \hbar \rabict \cos(\freqc t) \, \sigx + \hbar \rabis \cos(\freqs t + \phis) \, \sigz$, comprising constant Zeeman splitting \freql, time-dependent control by resonant drive at radiofrequency \freqc\, with swept amplitude \rabict, and a weak test signal parameterized by amplitude \rabis, oscillation frequency \freqs\, and phase \phis.
	We increase the control amplitude at constant sweep rate $\lambda$, realizing the continuously-varying Rabi frequency $\rabict=\Omega_\text{i} + \lambda t$.
	
	Transforming into the first rotating frame by $\hione=\hat{S}_1^\dagger\hlabt\hat{S}_1 - \iu \hbar \hat{S}_1^\dagger\,\hat{\dot{S}}_1$ where $\hat{S}_1 = \exp(\iu \freqc \sigz t / 2)$, and applying the rotating wave approximation (RWA) yields $\hone = (\hbar/2)\left(\detc\sigz + 2\rabis \cos(\freqs t + \phis)\sigz + \rabict\sigx\right)$.
	For convenience only, we now form $\honer$ by rotating $\hone$ through $-\pi/2$ around $\hat{y}$ so that, in the absence of both control errors ($\detc=0$) and signal ($\rabis=0$), the eigenstates are of $\sigz$, and are split by $\rabict$: this is our swept sensing qubit.
	
	\begin{figure}
		\includegraphics[scale=0.45]{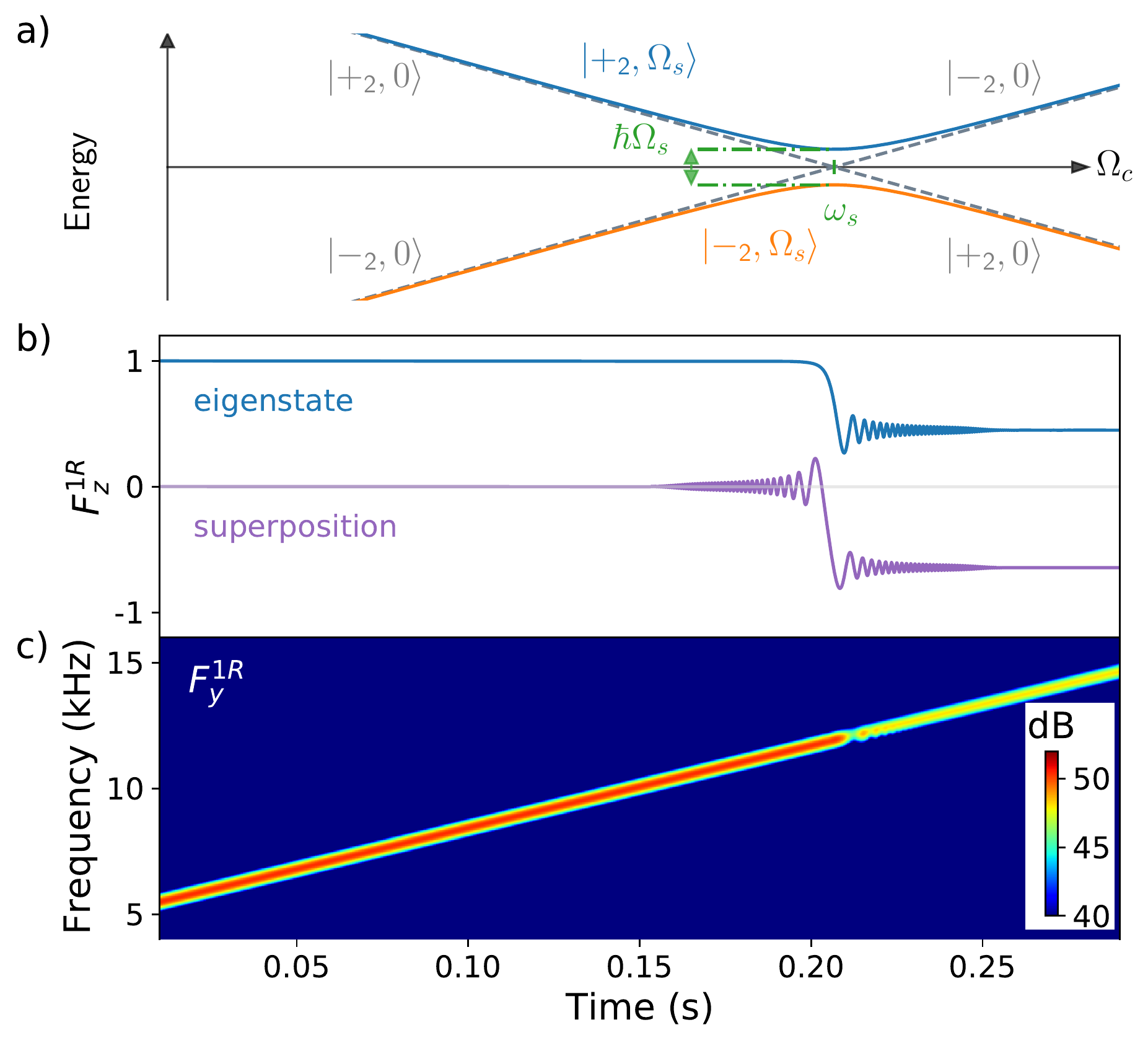}
		\caption{
			Non-adiabatic sensing: under a sweep of sensing frequency \rabict\, the location of, and gap between, adiabatic eigenstates $\ket{\pm_2,\rabis}$ in an avoided crossing (a) reveal frequency \freqs\, and amplitude \rabis\, respectively, of a sinusoidal signal coupling the bare states $\ket{\pm_2, 0}$. 
			Evolving spin projections (simulated, longitudinal (b), and transverse (c)) in the diabatic basis also encode signal parameters, but now as transition time, oscillations and asymptotic value, which all depend (b) on the initial state.
			Transverse spin projection, shown as spectrogram (c) for an initial superposition, oscillates at \rabict\, except near resonance where an avoided crossing emerges.
			A successful estimator robustly decodes signal parameters from measured spin projections.
		}
		\label{fig:avoided_crossings}
	\end{figure}
	
	If we suppose for a moment that the signal frequency \freqs\, is known, we may transform to a second frame co-rotating at \freqs\, and make a second RWA yielding
	$\htwo = (\hbar / 2) \left[ \left(\Omega_\text{i} + \lambda t - \freqs \right)\sigz+ \rabis \cos(\phis)\sigx - \rabis \sin(\phis)\sigy \right]$.
	This is the canonical Landau-Zener Hamiltonian, merely with a delayed resonance time $t_\text{res} = (\freqs - \Omega_i)/\lambda$, i.e. $\rabic(t_\text{res})=\freqs$.
	A signal of amplitude \rabis\, induces an avoided crossing in the instantaneous eigenstates of \htwo, as seen in Fig.~\ref{fig:avoided_crossings}(a).
	A sweep that commences from a dressed eigenstate remains exponentially close to it until just prior to $t_\text{res}$.
	For a weak signal, and hence a small gap, the crossing is non-adiabatic and a partial transition to the other diabatic eigenstate occurs, manifest as a change in the longitudinal spin projection \Fzoner\, (Fig.~\ref{fig:avoided_crossings}(b), blue).
	Henceforth we denote the time-dependent expectation value of spin projection on axis $i$ by $\spinproj{i}{P}=\spinproj{i}{P}(t)=\bra{\psi_P(t)}\hat{F}_i\ket{\psi_P(t)}$, where $\ket{\psi_P(t)}=c_{+}(t)\ket{+_P}+c_{-}(t)\ket{-_P}$ and $\ket{\pm_P}$ are the frame-$P$ eigenstates.
	$\hat{F}_i$ are the atomic hyperfine spin-1 operators, and as described below, at all times $F_i = 2\expval{\hat{\sigma}_i}$.
	Under a non-adiabatic transition through the avoided crossing, a clear amplitude drop in the oscillating transverse projection \Fyoner\, is evident in Fig.~\ref{fig:avoided_crossings}(c).
	Minimizing total residual $ || \spinproj{z\text{,meas}}{\text{1R}}(t) - \Fzoner(t, \freqs, \rabis, \phis) ||_2 $ between the continuously-measured spin projection and the Bloch equations solution yields signal parameters $\freqs,\,\rabis$~and~$\phis$ retrieved simultaneously from a single evolution.
	Probing the Landau-Zener transition structure in this manner subverts the conventional assumption in quantum sensing that the resonant frequency \freqs\, must be known a priori, or carefully determined, for sensitive measurement of amplitude. 
	
	We realize the protocol on ultracold rubidium-87 atoms in the $F=1$ hyperfine ground state.
	The ensemble of $1.8\times10^6$ atoms at $\unit[1.0]{\mu K}$ is approximately spherical with diameter $\unit[70]{\mu m}$, held in a crossed optical dipole trap.
	A static $\hat{z}$ bias field yields Zeeman splitting by $\freql = \unit[2\pi\times603.5]{kHz}$, with all atoms initially in the $F=1, m_F=-1$ state.
	We cancel the quadratic Zeeman shift of $\unit[53.5]{Hz}$ to well below $\unit[1]{Hz}$ with the ac Stark shift of a microwave field detuned from the clock transition by $\unit[387]{kHz}$, so that the Zeeman triplet is symmetrically split.
	The atomic density is sufficiently low that spin-mixing dynamics do not affect the transverse spin magnitude~\cite{liu_quantum_2009-1} appreciably during the sweep.
	With dephasing and interaction effects nulled, spin projections follow single-atom spin-$\half$ dynamics for at least $\unit[1]{s}$~\cite{jasperse_continuous_2017}.
	A resonant rf field ($\detc\approx0$) along $\hat{x}$ creates dressed states $\ket{\pm_1}$ split by \rabict\, tunable from $\unit[4]{kHz}$ to $\unit[50]{kHz}$.
	At low amplitudes, detuning errors \detc(t)\, generate increasingly large fluctuations in the splitting while at high amplitudes, Bloch-Siegert shifts require compensation.
	We add a test signal with weak amplitude~\rabis\, at frequency~\freqs\, to our bias along $\hat{z}$.
	This is the sinusoidal signal we will attempt to measure.
	
	We record a timeseries of the transverse spin projection, \Fxlab, with a minimally-destructive Faraday spin-light interface, tuned to the magic-zero wavelength of $\unit[790.03]{nm}$ and with photodetected power of $\unit[6]{mW}$ spatially mode-matched to the atoms~\cite{jasperse_continuous_2017}.
	Digitized at $\unit[5]{MSa/s}$, the spin projection is measured at the photon standard quantum limit (SQL). 
	Over the $\unit[300]{ms}$ measurement time, scattering of probe photons depletes the atom number by $\sim20\%$ and to account for this, we estimate the instantaneous total spin by a Hilbert transform method in a passband centered on \freql.
	Normalized by this amplitude, \Fxlab\, is proportional to the transverse spin per atom.
	
	This laboratory-frame measurement alternately probes the two transverse spin projections \Fxone\, and \Fyone\, of the first rotating frame as they precess past the laser propagation axis.
	The Faraday measurement is a quadrature-amplitude modulated signal
	$\Fxlab = \Fxone \cos(\freqc t) + \Fyone \sin(\freqc t)$, with carrier frequency \freqc\, and in-phase and quadrature amplitudes \Fxone\, and \Fyone, respectively.
	Numerical demodulation with complex local oscillator $\exp(\iu\freqc t + \phic)$ parallels the rotating-wave approximation, where the resulting frame-1 measurement is a complex baseband signal proportional to $\Fxone + \iu \Fyone$.
	In the sensing qubit frame~1R, this is $\Fzoner + \iu \Fyoner$: a continuous measurement of both the longitudinal, and one transverse, spin component of the evolving dressed state.
	See supplemental materials section II for further processing detail~\footnote
	{
		\label{suppmat} See Supplemental Material for detailed breakdown of apparatus (I), processing (II), analysis procedure (III) and preliminary multi-tone measurements (IV), which includes refs \cite{taylor_unambiguous_2023, starkey_scripted_2013, hilbert_grundzuge_1912, kornblith_juliadspdspjl_2022, savitzky_smoothing_1964, dierckx_curve_1995, bloch_nuclear_1946, ahmad_differential_2022, kelley_4_1999}.
	}.
	
	Much of the subtlety of quantum metrology lies in how we infer the measurand from the quantum measurement record.
	While inferring a Rabi frequency from measurements of Rabi flopping is routine, we now face the much more challenging inverse problem of inferring the Rabi frequency and radiation frequency of the weak signal from the Landau-Zener evolution of the spin.
	An ungainly analytic solution exists in terms of sums of products of parabolic cylinder functions~\cite{vitanov_landau-zener_1996-1}, however their Stokes structure appears to impede regression, and no analytic solution exists for multi-frequency signals. 
	
	Our solution was to retrieve signal parameters \rabis, \freqs\, and \phis\, by parameter estimation on the underlying system of coupled ordinary differential equations.
	Such parameter estimation is well-established for time-independent Hamiltonians~\cite{zhang_quantum_2014}, however for time-dependent evolution prior demonstrations employed repeated evolutions with varying time-independent control~\cite{de_clercq_estimation_2016}, varying initial state~\cite{siva_time-dependent_2023} or dependence on preceding measurements~\cite{naghiloo_achieving_2017}.
	Seeking a retrieval from a single unitary evolution, we regressed the numerical solution of the time-dependent Bloch equations to the data.
	We repeatedly solve the Bloch equations $\vec{\dot{F}}^\text{1R} = \vec{F}^\text{1R} \cross \vec{\Omega}^\text{1R}$, where $\vec{F}^\text{1R} = \Fxoner\hat{x}+\Fyoner\hat{y}+\Fzoner\hat{z}$  is the Bloch vector and $\vec{\Omega}^\text{1R} = 2\rabis \cos(\freqs t + \phis)\hat{x} + \rabict\hat{z}$ the Rabi vector, varying the signal parameters and computing the cost function as the $\ell^2$-norm of the residual between \Fzoner\, predicted by the Bloch solver~\cite{tsitouras_rungekutta_2011}, and the data. 
	Working in frame-1R admits direct extension to retrieving multiple sinusoids, but the cost function is highly oscillatory and finding the global minimum is challenging, even with only three parameters~%
	\footnote{
		We use the \texttt{OrdinaryDifferentialEquations.jl} ecosystem~\cite{rackauckas_differentialequationsjlperformant_2017} to identify the correct basin with a global optimizer~\cite{feldt_blackboxoptimjl_2018} before a fast local optimizer~\cite{mogensen_optim_2018} locates the minimum and extracts parameter covariances.
		The global optimizer is constrained only by requiring \freqs\, within the span, and positive \rabis\, less than a nominal maximum well into the adiabatic regime. See supplemental material \cite{Note1} section III.
	}.
	Solving the Bloch equations is exactly equivalent to solving the time-dependent Schr\"odinger equation for $\honer(t)$, but is faster and obviates calculating expectation values.
	
	While the retrieval uses only longitudinal spin data, successful estimation of parameters should result in the model agreeing with data in all three spin projections.
	Seeking to remove inessential dynamics independently of any particular solution, we transform to a spin-rotating frame-2S that co-rotates at the swept control Rabi frequency \rabict\, via $\hat{S}_\text{2S} = \exp( \iu \int_{0}^{t}\rabic(\tau)d\tau \, \sigz/ 2) = \exp( (\iu \Omega_\text{i}t/2 + \iu \lambda t^2/4) \sigz)$, as in Majorana's celebrated solution~\cite{majorana_atomi_1932,kofman_majoranas_2023}.
	Without making a RWA, the Hamiltonian becomes $\htwos=\hbar\rabis\cos(\freqs t+\phis)\left[\cos(\Omega_\text{i}t+\lambda t^2/2)\sigx-\sin(\Omega_\text{i}t+\lambda t^2/2)\sigy\right]$, which interestingly is multiplicative with the signal and a prospect for development of the inverse problem.
	After making the RWA, only low-frequency dynamics remain, and the trajectories take a compact and elegant form related to the Cornu spiral~\cite{zhuang_noise-resistant_2022}, unwinding towards resonance and curling again beyond~%
	\footnote{
		For a weak signal and evolving from an initial eigenstate, the Cornu spiral appears eidetically in orthographic projection with one focus centered on the pole; kinematically a large Bloch sphere rolls without slipping on the planar spiral~\cite{rojo_rolling_2010}.
	}. 
	
	We generate projections in this frame through further demodulation of the transverse spin projection \Fyoner\, with the swept complex carrier $\exp(\iu (\Omega_\text{i} + \lambda t/2)t +\iu\phi_{2S}(t))$, yielding \Fytwos\, and $-\Fxtwos$ as the real and imaginary components.
	These transverse frame-2S projections prove useful for improving retrieval precision via analysis of the spin azimuthal phase $\phi_{2S}(t) = \atantwo(\Fytwos,\Fxtwos)$, with residual slow winding of $\phi_{2S}$ attributable to small errors in \rabict.
	By fitting a smooth spline to the spin azimuth we can correct slow drifts in the Rabi control frequency, while leaving intact the small but rapid change in phase near resonance.
	While the optimizer considers only the longitudinal spin projection, these corrections enter the Bloch equations through \rabict, improving both the accuracy of the solution and its presentation in frame-2S.
	As $\hat{S}_\text{2S}$ commutes with $\hat{\sigma}_z$, the longitudinal spin projection is unchanged: $\Fztwos = \Fzoner$.
	This yields all three projections in this frame and so represents continuous state tomography of the time-evolving system by double demodulation.
	In this context, subsequent parameter retrieval under our protocol represents sensing by process tomography of an ineluctably time-dependent quantum process. 
	We now present a signal retrieval and its visualization on the Bloch sphere in frame-2S.
	
	\begin{figure*}
		\begin{center}
			\includegraphics[scale=0.65]{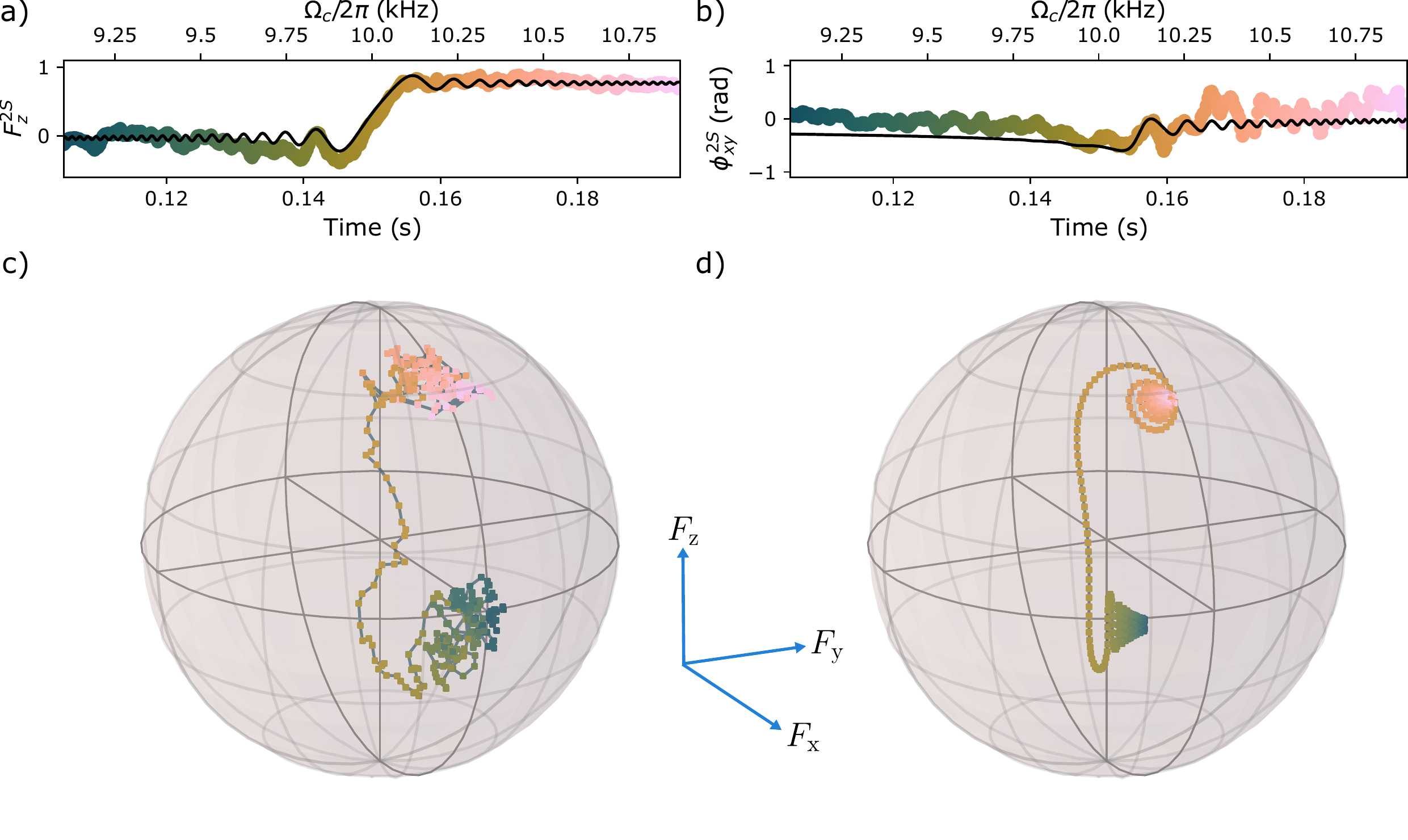}
			\caption{
				Fitting a quantum process model to a continuous measurement of unitary evolution estimates frequency and amplitude in a single shot.
				Timeseries for (a) longitudinal spin projection and (b) equatorial spin angle show the continuous measurement (color) and the model (black) retrieving precision parameter estimates of $\bs = \unit[3.34(2)]{nT}$, $\freqs = \unit[2\pi\times10.00434(16)]{kHz}$ and $\phis = \unit[3.93(14)]{rad}$.
				Evolution as Bloch sphere trajectories for smoothed data (c) and model (d) in frame-2S, with the same color mapping for time \cite{crameri_misuse_2020}.
			}
			\label{fig:bloch_plots_sup}
		\end{center}
	\end{figure*}
	
	The sensor is initialized in a superposition of dressed eigenstates $(\ket{+_\text{1R}} + \ket{-_\text{1R}})/\sqrt{2}$, equivalent to the lab-frame eigenstate $\ket{+_0}$.
	Rabi coupling is introduced, immediately commencing an upward sweep in amplitude from $\rabic=\unit[2\pi\times7]{kHz}$ to $\unit[13]{kHz}$ over $\unit[300]{ms}$.
	In the absence of a test signal, the adiabatic theorem indicates the amplitudes of the dressed states remain unchanged even as their splitting steadily increases.
	We introduce a weak test signal at $\freqs=2\pi\times\unit[10]{kHz}$, which drives a Landau-Zener transition evident in Fig.~\ref{fig:bloch_plots_sup}(a) (colored trace).
	Optimization in frame-1R converges after order $10^4$ iterations (Fig.~\ref{fig:bloch_plots_sup}(a), black trace), yielding a multiparameter estimate of $\bs = \rabis/\gamma = \unit[3.34(2)]{nT}$ for signal amplitude and $\freqs = 2\pi\times\unit[10.00434(16)]{kHz}$ for signal frequency.
	In fact, the digitally-synthesized sweep proceeds in steps of $2\pi\times\unit[3]{Hz}$, likely accounting for much of the frequency estimate residual (see Supplemental Material, section I~\cite{Note1}).
	
	Fig.~\ref{fig:bloch_plots_sup} shows data~(c) and model~(d) as trajectories on the Bloch sphere in frame-2S, with the common color code denoting time.
	With only longitudinal data used in parameter estimation, strong correlation between data and estimated solution of not only the longitudinal projection but of the full state evolution instills confidence parameter estimation has succeeded.
	A calibration shot with constant requested $\rabic = 2\pi\times\unit[10]{kHz}$ yielded an \emph{unswept} Rabi parameter estimate~\cite{white_unpublished_nodate} of $\bs=\unit[3.272(9)]{nT}$, covarying significantly with control error in \rabic\, the estimate of which varies by order \rabis\, between calibration shots. 
	This inadvertently revealed an inherent advantage in swept multiparameter estimation: while the resonance crossing may be shifted by such detuning error, the amplitude estimate remains almost entirely unaffected, making the swept sensor considerably more robust.
	
	As with the Rabi protocol, the initial state determines whether the Landau-Zener protocol is sensitive to phase.
	Measurements beginning with an initial superposition depend on the phase difference at resonance crossing.
	It is well-known that such phase-sensitive measurements have the highest possible (linear) sensitivity to radiation in quadrature~\cite{staudenmaier_phase-sensitive_2021}, while being completely insensitive to radiation in phase with the qubit.
	In contrast, the Rabi protocol beginning in an eigenstate is phase-insensitive, measuring purely amplitude, but at the price of being only quadratically sensitive to it.
	
	Our measurement presented in Fig.~\ref{fig:bloch_plots_sup} commenced with a superposition and thus should be phase-sensitive.
	The retrieved signal phase $\phis = \unit[3.93(14)]{rad}$ was, however, not repeatable which we attribute to the residual error in \rabic\, -- even after the spline interpolation correction -- integrated up over the $2\pi\times3000$ radian sweep prior to resonance.
	Repeatable amplitude retrieval despite this demonstrates a fundamental advantage of the Landau-Zener sensor: even for a signal exactly in phase at resonance, where a Rabi measurement would have zero sensitivity, the Landau-Zener time-series shows a characteristic transient providing frequency and amplitude information.
	Further, this suggests that amplitude and phase can be separately and unambiguously measured with linear sensitivity in a single initialization.
	We have also performed phase-insensitive retrieval of signal parameters commencing in a dressed eigenstate, retrieving frequency and amplitude without including a phase parameter. 
	
	Our retrieval variances correspond to  amplitude sensitivity of $\unit[11]{pT/\sqrt{Hz}}$, frequency sensitivity of $\unit[0.026]{Hz/Hz^{3/2}}$ and phase sensitivity of $\unit[0.084]{rad/\sqrt{Hz}}$.
	In general, the sensitivities appear to be nonlinear functions of sweep rate $\lambda$, sweep time, initial state, Faraday probe SNR and each other.
	Some insight follows from considering $\sqrt{\lambda}$, a critical frequency in swept-sine spectrum analyzers:
	in classical analyzers it defines the minimum undistorted RBW and hence spectral resolution~\cite{engelson_modern_1984}.
	In the quantum analyzer, $\sqrt{\lambda}$ is the adiabaticity threshold for \emph{amplitude}: for weak signals $\rabis < \sqrt{\lambda}$, the transition full-width approaches $\sqrt{\lambda/\pi}$ rather than $\rabis$~\cite{kofman_majoranas_2023,vitanov_transition_1999}.
	In this limit the quantum analyzer should equal the frequency resolution of its classical counterpart.
	Our sweeps of $\unit[6]{kHz}/\unit[300]{ms}$ are equivalent to RBW $(\sqrt{\lambda/\pi})/2 = \unit[100]{Hz}$, and so resolve 60 channels across the span.
	On average $6\times 10^3$ atoms scatter probe photons during the $\unit[5]{ms}$ channel transit~\cite{jasperse_continuous_2017}, giving a prospective atom SQL of $\unit[0.99]{pT/\sqrt{Hz}}$.
	Furthermore, under unitary evolution all time points after the transition are affected, implying a lower detection limit than this simple transit-time model predicts.

	A single-spin nanoscale analyzer~\cite{staudenmaier_phase-sensitive_2021} using Qdyne methods to retrieve single-tone signal parameters in a $\unit[1.5]{MHz}$ band around a fixed $\unit[1.51]{GHz}$ resonance was $\unit[74]{dB}$ less sensitive than our swept resonance analyzer, which uses $\unit[63]{dB}$ more spins. 
	Frequency and phase sensitivities were  almost identical.
	Sweeping the resonance not only covers  octaves of bandwidth but also obviates the dynamic range constraints of Rabi aliasing.
	
	Finally, we have not considered the quantum backaction of the continuous measurement.
	Simultaneous Faraday measurements of two spin quadratures furnished by Larmor precession yield planar spin squeezed states~\cite{colangelo_simultaneous_2017}, and it is plausible that analogous non-demolition backaction may enhance dressed-state spectral sensing.
	
	The linearly-swept qubit is an archetypal problem of time-dependent quantum mechanics, and the Landau-Zener solution thereof emerges naturally in quantum measurement of signals with frequency unknown.
	We have used this insight to demonstrate a Landau-Zener multiparameter estimator for frequency, amplitude and phase, operating on a single unitary evolution.
	The long coherence time of radiofrequency-dressed cold atoms enables a quantum spectrum analyzer achieving time-bandwidth product of 1800, amplitude precision of $\unit[20]{pT}$ and sub-Hertz resolution in frequency.
	We have retrieved rich spectra from multi-tone signals, with encouraging resolution of component frequencies, although so far without precision estimation of amplitudes~\cite{Note1}; we are optimistic that precision retrievals of rich spectra are possible.
	We envision applications including compact very-low-frequency (VLF) receivers, and microscale nuclear magnetic resonance (NMR) spectroscopy in Earth's field.
	More fundamentally, the analyzer heralds a new class of intrinsically time-dependent sensing protocols, neither Ramsey nor Rabi, which harness non-adiabaticity for quantum measurement.
	
	\begin{acknowledgments}
		We thank Russell Anderson, Michael Barson, James Saunderson and Alexander Wood for helpful discussions, and James Pollock and Maeva Berchon for preparing the apparatus.
		This work was supported by an Australian Government Research Training Program (RTP) scholarship, and funded by the Australian Research Council under Linkage Project LP200100082.
	\end{acknowledgments}

	%

	
	\clearpage
	\onecolumngrid
	\begin{center}
		\textbf{\large Supplemental Materials: Quantum spectral analysis by continuous measurement of Landau-Zener transitions}
	\end{center}
	\setcounter{equation}{0}
	\setcounter{figure}{0}
	\setcounter{table}{0}
	\setcounter{section}{0}
	\renewcommand{\thesection}{S-\Roman{section}}
	\setcounter{page}{1}
	\makeatletter
	\renewcommand{\theequation}{S\arabic{equation}}
	\renewcommand{\thefigure}{S\arabic{figure}}
	
		\section{I: Apparatus}
	
	\subsection{Preparation of the sensing qubit}
	Our sensing qubit comprises $1.8\times10^6$ rubidium-87 atoms, produced by a conventional laser cooling and evaporative cooling sequence, and held in a crossed optical dipole trap at $\unit[1.0]{\mu K}$, forming an approximately spherical thermal cloud $\unit[70]{\mu m}$ in diameter.  
	
	While further evaporative cooling can take our system to degeneracy, producing a spinor Bose-Einstein condensate (BEC), here we work instead with a larger sample of thermal atoms just above the BEC transition.
	All partial waves above s-wave are frozen out and the only operating collision channels are either entirely elastic spin-preserving collisions, which do not affect the spin evolution, or spin-mixing collisions within the $F=1$ hyperfine manifold.
	At densities significantly lower than the degenerate Bose gas, these spin-mixing collisions proceed sufficiently slowly as to negligibly affect the transverse spin magnitude, in comparison to residual atom-loss due to off-resonant scattering of the Faraday probe.
	In the absence of the Faraday probe, the trap lifetime exceeds $\unit[20]{s}$, limited by scattering of dipole trap light, and background gas collisions.
	With the Faraday probe beam on, off-resonant scattering reduces the lifetime to $\unit[1.3]{s}$.
	Residual magnetic field and vector light-shift gradients reduce the Zeeman coherence time to around $\unit[0.5]{s}$. 
	
	A static $\unit[86.1]{\mu T}$ bias field along $\hat{z}$ produced by a Helmholtz coil pair creates a Larmor resonance at $\freql=2\pi\times\unit[604.5]{kHz}$.
	Interference fields from the AC line at $\unit[50]{Hz}$, and its odd harmonics, induce fluctuations of order $\unit[2]{kHz}$ about $\freql$. 
	These fluctuations are very stationary, and by triggering the experiment to a zero-crossing of the AC line voltage, we are able to suppress these fluctuations by up to $\unit[30]{dB}$ through feed-forward techniques~\cite{taylor_unambiguous_2023}:
	a single calibration shot determines the Fourier coefficients of the line harmonics at the atomic sensor, which are fed forward in anti-phase in future shots.
	
	At our bias field, the ground-state Zeeman triplet is not quite equally split; with the $m_F=0$ central level shifted down by $\unit[53.5]{Hz}$ with respect to the midpoint of the $m_F=-1$ and $m_F=+1$ levels.
	This quadratic Zeeman shift would lead to collapse and revival of the transverse spin.
	We cancel the shift by inducing an equal and opposite ac Zeeman shift by applying a stationary microwave field detuned from the $F=1, m_F=0 \leftrightarrow F=2,m_F=0$ clock transition by $\unit[387]{kHz}$.
	With the quadratic term removed from the Hamiltonian, the three $F=1$ energy levels are split evenly and can be treated as a degenerate pair of two-level systems under the Majorana representation \cite{majorana_atomi_1932}.
	
	Atoms are initialized in the $F=1, m_F=-1$ weak-field seeking state during the magnetic trap stage, and remain in this state through evaporation in the optical dipole trap.
	To prepare the dressed sensing qubit, we turn on resonant ($\detc\approxeq0$) radiofrequency coupling transverse to our bias field.
	While this cw control field has fixed frequency for the duration of the experiment, achieving a linear sweep in dressed state splitting requires a precisely linear sweep in the rf amplitude.
	We realize this by stepping a A direct digital synthesizer (DDS) through a smaller, multi-turn coil.
	Initialized from a PulseBlaster direct digital synthesis (DDS) output, the signal is amplified with a MiniCircuits LZY-22+ $\unit[30]{W}$ rf power amplifier, attenuated to feed $~\unit[34]{dBm}$ of rf power to the coil and provide a corresponding maximum Rabi frequency of $\rabic^\text{max}\approxeq2\pi\times\unit[50]{kHz}$.
	Higher frequencies of up to $\unit[200]{kHz}$ were applicable without attenuation, but maximum range was sacrificed for improved precision in the desired low-kHz band.
	Taking advantage of our source being a DDS, amplitude changes may be driven during the experiment, cycling through a range of up to $2000$ points.
	This amplitude ramp is set as a calibrated polynomial, such that the amplified rf power delivered by the coil to the atoms is as close to a linear ramp as possible to match our modeling for the Landau-Zener Hamiltonian.
	Small remaining errors may be corrected in post-processing, detailed in section II.
	
	\subsection{Generation of test signal}
\begin{figure}
		\includegraphics[scale=0.8]{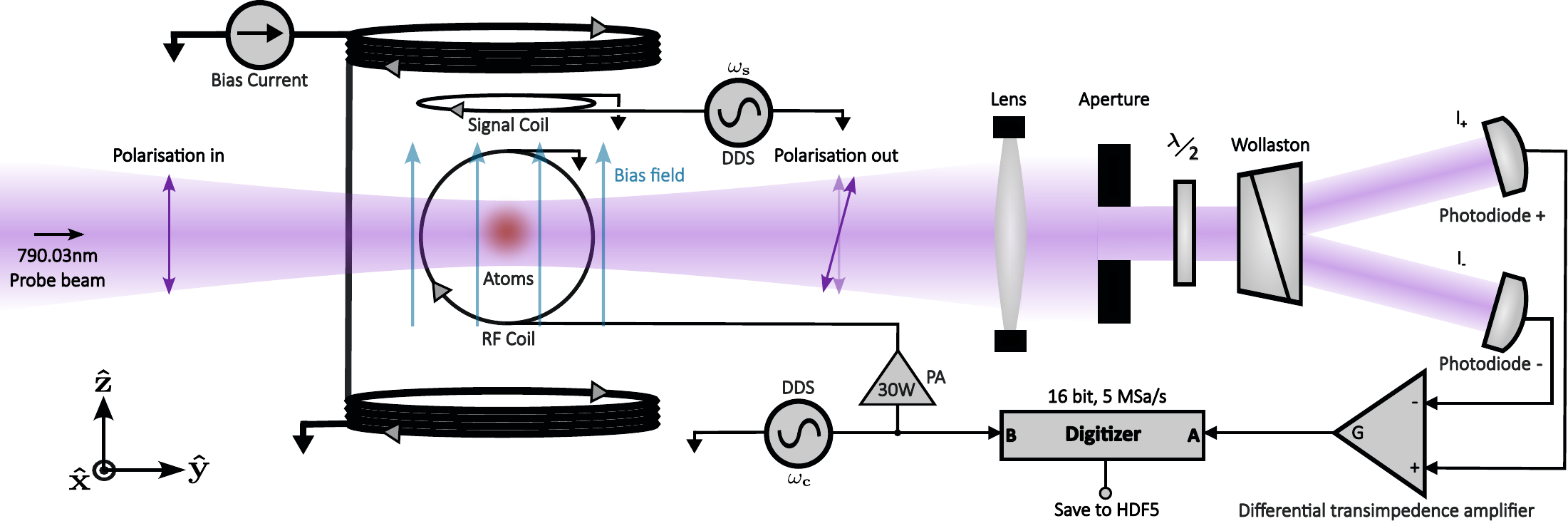}
		\caption{Block diagram of key apparatus components during science stage of the experiment.
			A crossed-beam dipole trap holds the ultracold atomic ensemble of $1.8\times10^6$ atoms at $\unit[1]{\mu K}$, with the Faraday probe beam traveling along one of the beamlines.
			Polarization of incident probe light is rotated through the optical Faraday effect dependent on the evolving spin state of the ensemble, with this change recorded via a balanced polarimeter, consisting of a half waveplate, Wollaston prism, photodiode pair and differential transimpedance amplifier in series.
			Recorded polarimeter signal is digitized at \unit[16]{bit}, \unit[5]{Msps} along with a copy of the radiofrequency reference.
			Key magnetic sources generating the DC bias field, magnetic test signal and radiofrequency dressing field are highlighted in their respective relative axes, with a coordinate legend in the bottom right.}
		\label{fig:supp_apparatus}
	\end{figure}

	Along this same axis, a small, single turn coil is added for the specific purpose of injecting test signals produced from an amplified input DDS, as seen in Fig.~\ref{fig:supp_apparatus}.
	The use of a small coil for this purpose improves resolution when opting for weak amplitudes, with older measurements instead injecting signals using the Helmholtz coil pairs as an additional modulation on the static bias field.
	Through this, signals are able to be introduced at much lower amplitudes than before, with order \unit{nT} achievable without aliasing.
	This surpasses the Helmholtz coil limit of $\approxeq\unit[15]{nT}$, roughly corresponding with the smallest voltage step resolvable on the Helmholtz coil source on top of the static bias.
	
	\subsection{Continuous transverse spin measurement}
	The decision to perform this experiment with an ultracold ensemble rather than a BEC stems from the Faraday measurements we perform.
	Pictured in Fig.~\ref{fig:supp_apparatus}, an off-resonant probe beam tuned to the magic-zero wavelength of $\unit[790.03]{nm}$~\cite{jasperse_continuous_2017} performs weak measurements of the transverse spin via the Faraday spin-light interface.
	Probe photons are manipulated such that their polarization is linear upon interaction with the atoms, where a polarization rotation is imparted via the optical Faraday effect dependent on the expected quantum state of the ensemble.
	It was found that with our current setup, improved Faraday signal arising from an order of magnitude increase in trapped atoms proved more beneficial than the reduction in atom shot-noise for a BEC state.
	A total photodetected power of $\unit[6]{mW}$ is measured on the balanced polarimeter comprising half-waveplate, Wollaston prism, series-connected reverse-biased photodiode pair and transimpedance amplifier, as shown in Fig.~\ref{fig:supp_apparatus}. 
	Rotation of the probe beam polarization by interaction with the atomic spin is encoded in the bipolar output voltage of the polarimeter, which measures in a bandwidth exceeding $\unit[1]{MHz}$ with electronic noise floor, dominated by the transimpedance Johnson noise, approximately \unit[10]{dB} below the photon shot noise. This delivers wideband polarimetry at the photon standard quantum limit (SQL).
	
	The polarimeter signal is digitized at 16 bits and $\unit[5]{MSa/s}$ on channel A of an AlazarTech ATS9462 digital acquisition (DAQ) card.
	On Channel B we record a copy of the rf drive signal.
	This provides an rf reference trace used during signal processing for demodulation.
	Both channels are saved to a unique HDF5 file per shot~\cite{starkey_scripted_2013}, and the digital signal processing described in Section II below is performed offline. 
	
	\subsection{Time-frequency structure of the Faraday signal}
	For calibration purposes, the Faraday recording extends beyond the swept measurement protocol bounds.
	Before the sweep begins, we have two calibration phases to assist with characterizing noise limits; a short, \unit[10]{ms} wait time is recorded with no Faraday probe present to record electronic noise. The Faraday probe is then turned on and acquisition continues for a further \unit[5]{ms} prior to the rf pulse that initiates Larmor precession. During this period, photon shot noise is present without any atomic signal. 
	
	Initial state preparation is then performed if necessary and the protocol begins.
	For the data presented in this manuscript our initial state for the protocol is the frame-0 (laboratory frame) eigenstate $F=1, m_F=-1$.
	We then turn on resonant radiofrequency coupling at \freqc, which induces continuous Rabi flopping at \rabic, or equivalently, dresses the bare states.
	As the spin Rabi flops between laboratory eigenstates, we observe full-depth amplitude modulation of the carrier at \freqc: the carrier disappears and we instead see two sidebands split by $2\rabict$, as seen in the spectrogram in Fig.~\ref{fig:sup_processing}(b).
	We immediately commence the up-sweep of \rabic\, by linearly increasing the rf amplitude. This is visible in the spectrogram as the sidebands diverging linearly.
	As the sensing qubit splitting crosses resonance with the test signal at \unit[10]{kHz}, the Landau-Zener transition occurs, observed as a partial transfer of amplitude from the sidebands into the carrier.
	While this is sufficient to detect a transition, we are able to determine signal parameters much more precisely, by working with demodulated quadratures extracted from the Faraday signal, as described below.
	
	\section{II: Signal processing to extract spin projections}
	
\begin{figure}
		\includegraphics[scale=1.06]{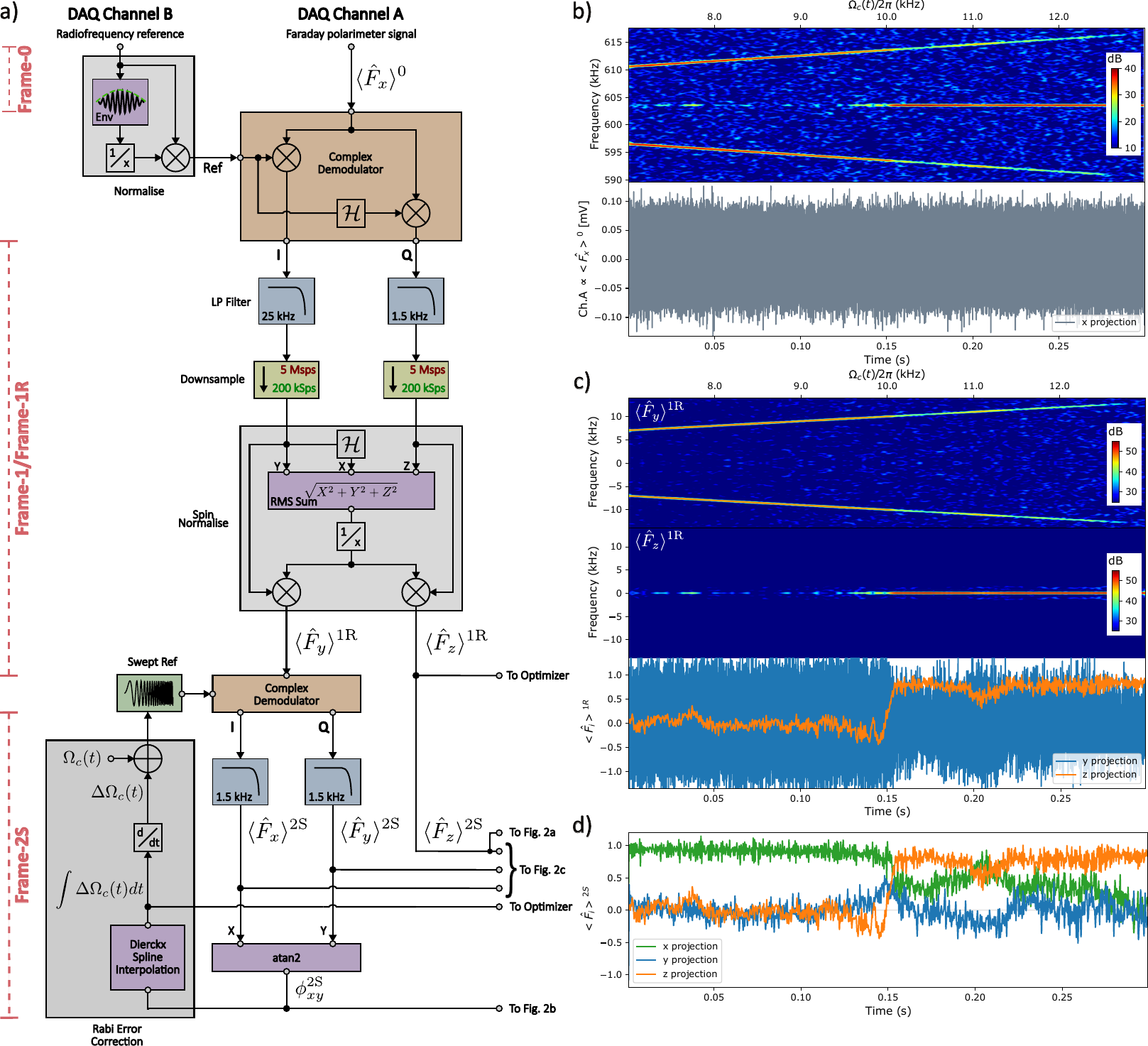}
		\caption{(a) Flow chart representation of the complete pre-processing pipeline to prepare our recorded frame-0 measurements for multiparameter estimation.
			Raw frame-0 data undergoes complex demodulation with the recorded reference trace into frame-1/1R, requiring filtering, downsampling and normalization of the Bloch vector to yield spin-normalized projections \Fyoner\, and \Fzoner.
			The transverse Frame-2S projections are retrieved via a second complex demodulation process via the swept carrier \rabict.
			Rabi error correction proves necessary to remove remaining phase evolution from Rabi frequency error, with the resulting frequency correction $\Delta\rabict$ both passed to the optimizer to improve retrieval precision, and to redo complex demodulation to isolate only true Frame-2S dynamics.
			All processed components used for optimization or figures are clearly signposted with output terminals and labels.
			(b) The recorded Frame-0 data presented as both spectrogram (top) and timeseries (bottom), showing that while clear frequency information is encoded with the carrier band at \freql\, and frequency-modulated sidebands separated by $\pm\rabict$, the timeseries is too polluted with noise for parameter estimation.
			(c) Frame-1R spectrograms for the transverse \Fyoner\, (top) and longitudinal \Fzoner\, (middle) timeseries shows two key points; demodulation has removed underlying oscillation at \freql, while separating low-frequency transition dynamics from oscillations at \rabict.
			The timeseries (bottom) show a clear Landau-Zener transition in the processed data trace on which parameter retrieval is performed (orange).
			(d) The final Frame-2S data traces presented on the Bloch sphere in Fig.\ref{fig:bloch_plots_sup}(c), with spectrogram unnecessary to prove only low-frequency dynamics remain.}
		\label{fig:sup_processing}
	\end{figure}

	The quantum sensing protocol presented in this work requires an extensive analysis pipeline for extracting rotating-frame spin projections.
	Here, we provide additional detail on these steps. Fig.~\ref{fig:sup_processing}(a) shows the process in block diagram form. 
	
	\subsection{Continuous quantum state tomography by successive demodulation: First stage}
	Before parameter retrieval can be performed, the recorded data must go through a sequence of pre-processing to extract from the recorded frame-0 spin projection the more useful frame-1R spin projections.
	Our raw measurements, shown in Fig.~\ref{fig:sup_processing}(b) in both timeseries and spectrogram form, cover a wide bandwidth intercepting considerable photon shot noise, and exhibit the expected exponential decay in total signal power due to atom loss.
	We begin by removing out-of-band noise with a bandpass filter applied about \freqc\, with a width greater than double the final Rabi frequency $2\Omega_\text{f}$, such that all Rabi modulation lies within the allowed band.
	
	To deal with the layered modulation, we demodulate the raw signal using the reference rf trace recorded on channel B of the DAQ.
	The rf reference is converted into a complex carrier via the Hilbert transform~$\mathcal{H}$~\cite{hilbert_grundzuge_1912}, which takes a real timeseries and induces a pan-spectral $\pi/2$ phase shift, forming the analytic signal $V_\text{ref} + \iu \mathcal{H}\left[ V_\text{ref} \right]$.
	Normalizing this analytic signal by its instantaneous modulus removes the linear amplitude ramp. Overall, this reference channel processing converts the digitized real signal $V(t)\cdot \cos(\freqc t)$ into $1\cdot\exp( \iu \freqc t)$.
	Additionally, a fixed phase correction of $65.0\deg$ is applied to the analytic reference, accounting for the fixed electronic, atomic and electro-optical phase shifts that occur in the apparatus, between rf generation at the DDS and eventual digitization. 
	
	We then extract the two transverse spin projections in frame-1R by complex demodulation. 
	Taking the numerical product of the Faraday signal, proportional to transverse spin projection $\Fxlab = \Fzoner\cos(\freqc t)+\Fyoner\sin(\freqc t)$, with the complex carrier $\exp(\iu\freqc t)$, and low-pass filtering realizes a complex demodulator, with output real and imaginary components proportional to \Fzoner\, and \Fyoner, respectively:
	\begin{align}
		\Fxlab\cdot\exp(\iu\freqc t) = \frac{\Fzoner}{2}(1+\exp(2\rabic t))+\iu\frac{\Fyoner}{2}(1-\exp(2\freqc t)).
	\end{align}
	Butterworth FIR low-pass (LP) filters with cutoff frequencies of \unit[1.5]{kHz} (well below $\Omega_\text{i}$), and \unit[22]{kHz} (just above $\Omega_\text{f}$), are applied to the real and imaginary outputs, respectively, removing terms at $2\rabic$.
	As \Fzoner\, consists purely of low-frequency components (frequencies well below \rabic), a lower cutoff was used for the real output to further reduce the noise bandwidth.
	With this, we find our frame-1R spin projections as
	\begin{align}
		\Fzoner &= 2\cdot\text{LP}_\text{1.5 kHz}\Big[\text{Re}\left[\Fxlab\cdot\exp(\iu\freqc t)\right]\Big] \, \text{and} \\
		\Fyoner &= 2\cdot\text{LP}_\text{22 kHz}\Big[\text{Im}\left[\Fxlab\cdot\exp(\iu\freqc t)\right]\Big].
	\end{align}
	Since the highest frequencies involved are considerably lower than before with removal of modulation at \freql, we also downsample from $\unit[5]{MSa/s}$ to $\unit[200]{kS/s}$ using the resample function in Julia's DSP package~\cite{kornblith_juliadspdspjl_2022}.
	These frame-1R filtered projections are shown in Fig.~\ref{fig:sup_processing}(c).
	Of note is how, after demodulation, Rabi information has been isolated in the \Fyoner\, projection, evident in the two spectrograms in Fig.~\ref{fig:sup_processing}(c).
	
	\subsection{Faraday signal normalization}
	These signals $F_i^\text{1R}(t)$ are strictly proportional to spin projections, but with proportionality constant depending on, inter alia, the number of atoms $N(t)$, which decays over time. 
	However, we know that as the quadratic Zeeman effect is canceled, the norm of the single-atom spin projections (i.e. the length of the Bloch vector) is $F\hbar$, and so writing the single-atom Bloch vector as $\vec{F}_s(t)=(F_x(t),F_y(t),F_z(t))$, we can relate the length of the \emph{measured} Bloch vector $\vec{F}_\text{meas}(t)$ to it by 
	\begin{equation}
		\left|\vec{F}_\text{meas}(t)\right| = \left|\alpha N(t) \vec{F}_s(t)\right| = \left|\alpha N(t)\right|\vec{F}_s(t) \, ,
	\end{equation}
	where $\alpha$ unifies the constants. 
	Normalization of our measured projections can therefore be achieved by computing the vector magnitude of our recordings, $\sqrt{F_x(t)^2 + F_y(t)^2 + F_z(t)^2}$ equivalent to $\left|\alpha N(t)\right|$, and dividing through by this time-dependent envelope.
	Doing so with our frame-1R projections, noting we can estimate the missing projection as $\Fxoner\approxeq\mathcal{H}\left[\Fyoner\right]$, accounts for all mechanism which homogeneously attenuate spin polarization including atom loss and dephasing. 
	A Savitzky-Golay filter~\cite{savitzky_smoothing_1964} smooths the envelope estimate so that only relatively slow attenuating processes are considered, and the data is normalized per-point by this envelope, leaving a unit Bloch vector at all times.
	
	\subsection{Continuous state tomography: second demodulation stage}
	For visualization purposes, we choose to perform a second demodulation from frame-1R into the swept second frame, frame-2S, which tracks the slewing Rabi frequency \rabict.
	To perform numerical demodulation into this frame, a complex carrier is introduced with the swept Rabi \emph{phase} $\int_0^{t}\rabic(\tau)\,d\tau$.
	Multiplying this complex carrier $1\cdot _\text{car}^\text{2S} = \exp(\iu\left(\Omega_\text{i} + \lambda t/2\right)t)$ with the transverse frame-1R projection
	\begin{equation}
		\Fyoner = \Fytwos\cos((\Omega_\text{i} + \lambda t/2)t) - \Fxtwos\sin((\Omega_\text{i} + \lambda t/2)t) \, ,
	\end{equation}
	and then LP filtering below \unit[1.5]{kHz} yields the frame-2S projections $-\Fxtwos$, and \Fytwos, as the real and imaginary components, respectively.
	Because $\Fzoner\equiv\Fztwos$ this resolves all three spin projections in frame-2S, and realizes fully time-dependent quantum state tomography under our protocol in this frame.
	The frame-2S projections are shown as time-series in Fig.~\ref{fig:sup_processing}(d), and as Bloch sphere trajectories in Fig.~2(c).
	
	\subsection{Rabi control error correction}
	While frame-2S has benefits for visualization, it also proves useful for the purpose of error correction.
	The complex carrier employed to demodulate into frame-2S assumes a perfectly linear sweep from $\Omega_\text{i}$ to $\Omega_\text{f}$.
	We correct for most of the slight gain nonlinearity of our LZY-22+ rf power amplifier by predistorting the amplitude synthesized by the DDS, but even the small residual errors, at the precision of our sensor, affect parameter estimation.
	To determine this residual error, we take advantage of the fact that the spin is stationary in frame-2S except in the immediate vicinity of the LZ transition.
	Any slow winding of the Rabi phase remaining is entirely attributable to deviation of the Rabi frequency from the linear sweep prescribed.
	
	We extract the azimuthal phase angle as $\phixy = \atantwo(\Fytwos,\Fxtwos)$.
	To retrieve the remnant Rabi phase error, we interpolate a one-dimensional cubic spline using the Dierckx spline fitting library~\cite{dierckx_curve_1995} implemented as \texttt{Dierckx.jl}.
	A smoothness parameter of $200$ captures the residual Rabi drift, $\Delta\phixy$, while excluding the brief Landau-Zener transition period.
	
	With the remnant Rabi phase modeled, two things become possible.
	First, we may perform demodulation into frame-2S a second time, this time with the error correction passed into the swept complex carrier function.
	This isolates the Landau-Zener evolution for visualization. 
	Second, the corrected Rabi frequency timeseries informs the retrieval in frame-1R, improving the precision of multiparameter estimation, making our protocol self-correcting for control amplitude errors.
	
	\section{III: Spectral analysis by multi-parameter estimation}
	
	\subsection{The forward problem: Bloch equations}
\begin{figure}
		\includegraphics[scale=0.9]{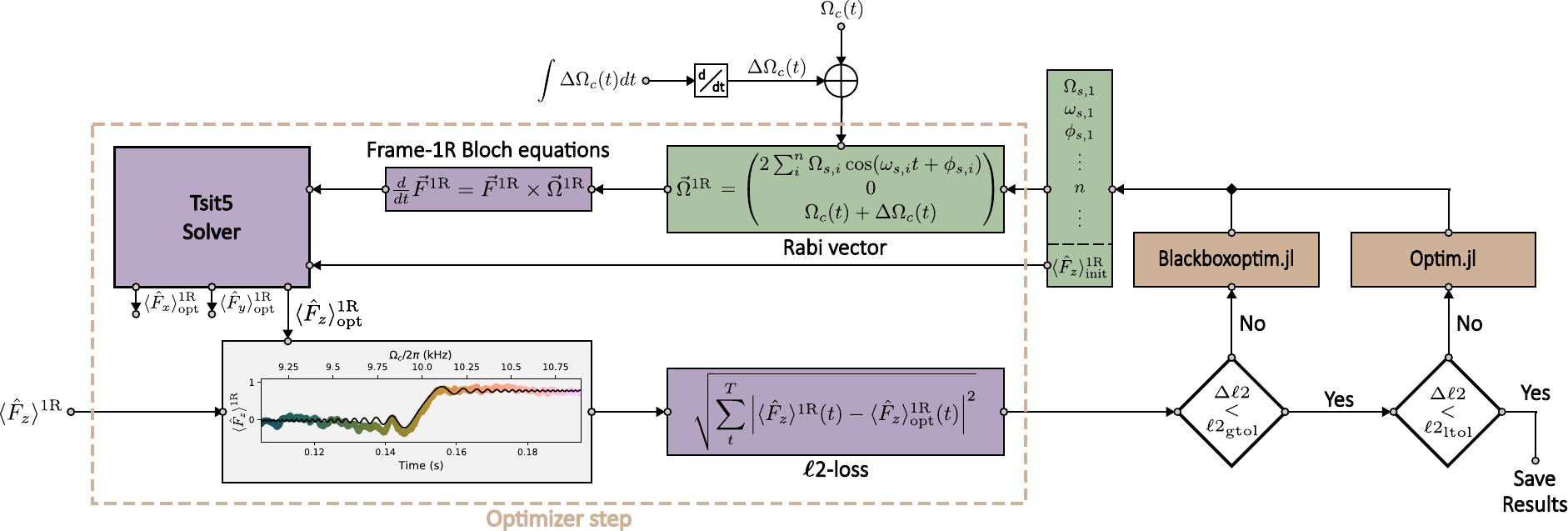}
		\caption{Block diagram presenting our optimization pipeline, performing inverse problem solving of the frame-1R Bloch ODEs and minimizing the $\ell2$-norm between input \Fzoner\, from the DSP pipeline.
			Each step in the loop computes $\spinproj{z,\text{opt}}{\text{1R}}$, taking signal parameters and initial $F_z$ projection as variables from the optimizer.
			First loops are performed through a global optimizer in \texttt{BlackboxOptim.jl}, until a global tolerance for minimum change in $\ell2$-norm is exceeded.
			A secondary loop in a local optimizer implemented in the \texttt{Optim.jl} package is then performed on the global result to find the local minimum.}
		\label{fig:supp_optim}
	\end{figure}

	The Bloch equations \cite{bloch_nuclear_1946} in frame-1R are  
	\begin{align}
		\dv{\Fxoner}{t} &= \Omega_z^\text{1R}\Fyoner - \Omega_y^\text{1R}\Fzoner \, , \\
		\dv{\Fyoner}{t} &= -\Omega_z^\text{1R}\Fxoner + \Omega_x^\text{1R}\Fzoner \, , \\
		\dv{\Fzoner}{t} &= \Omega_y^\text{1R}\Fxoner - \Omega_x^\text{1R}\Fyoner \, .
	\end{align}
	Referring to our frame-1R Hamiltonian, we can decompose it into the Rabi vector components $\left(\Omega_x^\text{1R},\Omega_y^\text{1R},\Omega_z^\text{1R}\right) = \left(2\rabis\cos(\freqs t + \phis), 0, -(\Omega_\text{i}+\lambda t+\Delta\phixy)\right)$, including the Rabi error correction.
	The forward solver is defined via the \texttt{DifferentialEquations.jl} package~\cite{rackauckas_differentialequationsjlperformant_2017}, evaluated using the Tsit5 modified Runge-Kutta method~\cite{tsitouras_rungekutta_2011} with a timestep of $\unit[8]{\mu s}$.
	Through iterative optimization of the forward solver in an optimization loop depicted in Fig.~\ref{fig:supp_optim}, we may perform multiparameter estimation on the unknown signal without requiring assumptions, representing a general solution to the problem our protocol poses.
	
	\subsection{Parameter estimation by global optimization}
	
	For any parameter estimator to function, it needs some measure of how close the estimated solution is to the recorded data.
	To this end, our cost function is least-squares: the $\ell2$-norm of the difference between the data timeseries of \Fzoner, and the Bloch solver solution.
	The objective is to minimize this cost function to its global minimum.
	This cost function uses only the \Fzoner\, spin projection.
	It is possible to implement a cost function which uses, for example, both \Fyoner\, and \Fzone, which would seem to use most if not all available data.
	We have found that with the narrower noise bandwidth of \Fzoner\, solution convergence was both quicker and more consistent in comparison to when supplemented with \Fyoner.
	Of note however is that we have used some of the information in \Fyoner\, in performing Rabi error correction described above.
	Our cost function takes four parameters: the amplitude \rabis, frequency \freqs\, and phase \phis\, of the signal, and the projection of the initial state \Fzoneri.
	This final parameter accounts for errors in the initial state preparation.
	
	\subsection{Global optimization stage}
	
	With the problem defined, we perform the first phase of our estimation with an unbiased global estimator, implemented using the \texttt{BlackBoxOptim.jl} package~\cite{feldt_blackboxoptimjl_2018}.
	We adopt the default optimization algorithm for BlackBoxOptim, an adaptive differential evolution \texttt{DE/rand/1/bin} algorithm with radius limited sampling.
	The target solution is mutated randomly within an adaptive radius and compared against the current optimal solution~\cite{ahmad_differential_2022}.
	A global selection process dictates the superior solution, in this case by comparing the cost function value, and keeps the best one, converging towards the optimal solution.
	The adaptive version of this algorithm we use selects trial values biased towards the current target, improving the search efficiency.
	
	To keep optimization free from bias, we try to constrain the optimizer as little as possible and do not pass it an initial guess.
	We bound search parameters with the sweep range for frequency $\freqs\in\left[\Omega_\text{i}=2\pi\times\unit[7]{kHz}, \Omega_\text{f}=2\pi\times\unit[13]{kHz}\right]$,  phase to lie within a single cycle $\phis\in\left[0,\unit[2\pi]{rad}\right]$, and initial \Fzoner\, unconstrained ($\Fzoneri\in\left[-1,1\right]$).
	For amplitude, we set broad bounds with a lower limit much weaker than the applied signal amplitude and with the upper limit well into the adiabatic regime: $\rabis\in\left[2\pi\times\unit[5]{Hz},2\pi\times\unit[300]{Hz}\right]$.
	A calibrated minimum change in cost function dependent on noise background acts as the stopping condition.
	This leads to typically $\leq10^4$ steps taking $\leq\unit[10]{mins}$ on an single Intel i7-10850H core.
	
	\subsection{Local optimization stage}
	The best result from the global optimizer is saved, along with information about the stopping condition and runtime, before being passed into a local optimizer to complete the process.
	We implement the Broyden-Fletcher-Goldfarb-Shanno (BFGS) local optimization algorithm~\cite{kelley_4_1999} via the \texttt{Optim.jl} package~\cite{mogensen_optim_2018}.
	Taking the global optimizer result as the initial condition, the local optimizer performs gradient descent to find the local minimum.
	This is much more rapid, typically taking only a few steps over $<\unit[5]{s}$ to determine the final value, and the surrounding gradients and curvatures.
	This optimizer returns an inverse Hessian matrix of covariances; the parameter uncertainties are the square-roots of the variances on the diagonal.
	These results are the final retrieved parameter values cited in-text with associated uncertainties, while the corresponding ultimate solution of the forward problem furnishes the frame-1R evolution time-series. 
	
	To take these final estimated spin projections and plot on the frame-2S Bloch sphere as seen in Fig.~2(d), the same swept demodulation (with corrected Rabi error $\Delta\phixy$) process we perform on our data is repeated with the frame-1R optimization results.
	
	\section{Fit precision, calibration and retrieval error}
	
	One of the major benefits of our spectrum analysis protocol is that it measures signal parameters \emph{independent of detuning}, due to its swept nature.
	While frequency is well-defined by the DDS producing the signal fed into the coil and phase similarly well defined, amplitude is a trickier beast.
	
	\subsection{Test signal amplitude calibration}
\begin{figure}
		\includegraphics[scale=0.6]{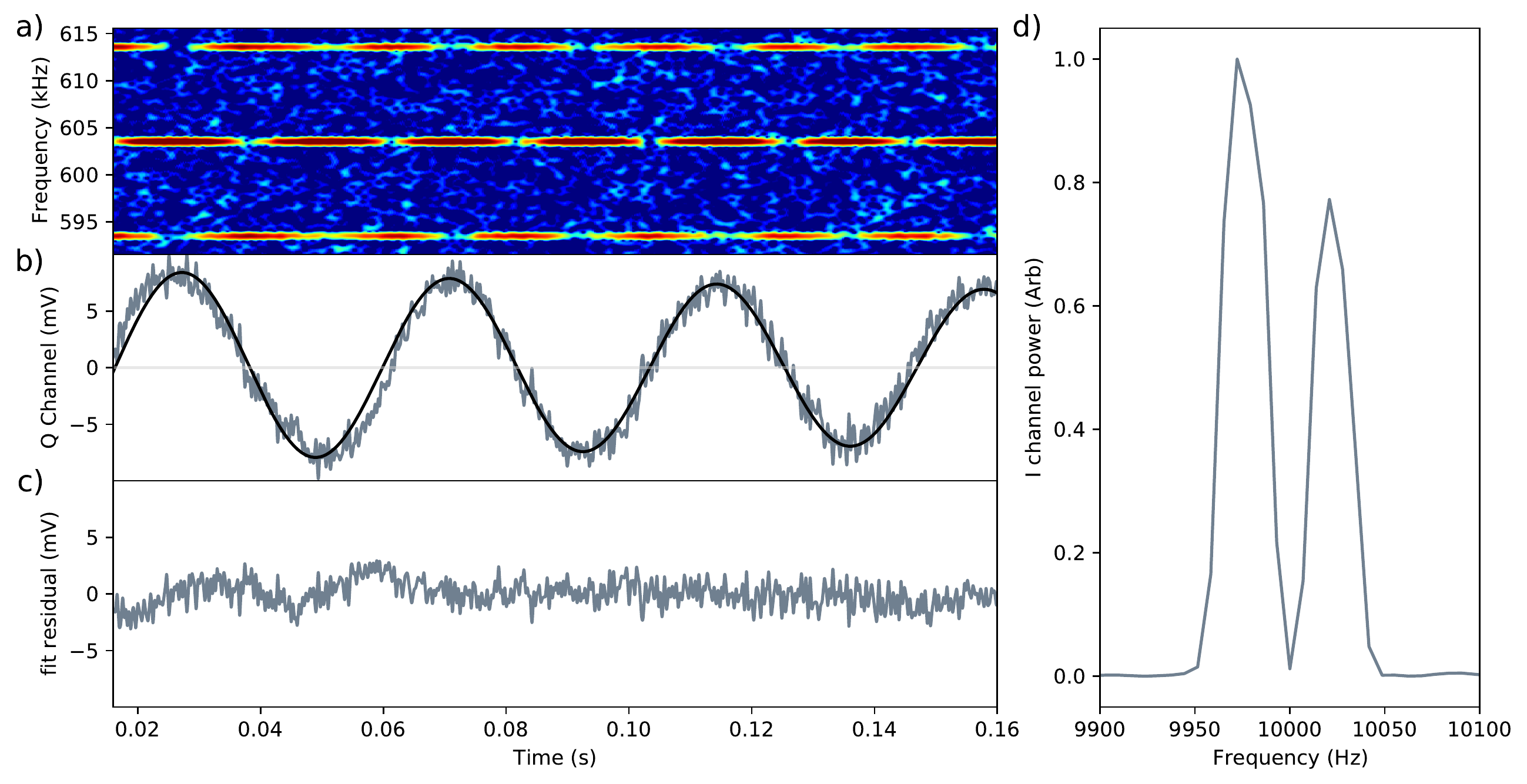}
		\caption{Frame-0 spectrogram (a), demodulated Q channel proportional to \Fyoner\, with a damped sinusoidal fit (b), and the fit residual (bottom left), for a resonant Rabi calibration shot of the signal measured with the LZ spectrum analyzer presented in the manuscript.
			Retrieved total Rabi amplitude of $\Omega_\text{s,R}^\text{cal} = \unit[22.97(1)]{Hz}$ was found as the fm demodulation frequency, before estimated detuning of $\Delta_\text{s}=-2\pi\times\unit[2.1(6)]{Hz}$ from the upper and lower \Fyoner\, sideband amplitudes (d) yields a corrected estimate of $B_\text{s}^\text{cal} = \unit[3.272(9)]{nT}$.}
		\label{fig:supp_calib}
	\end{figure}

	We perform calibrations via \emph{unswept} Rabi measurements read out with the Faraday probe.
	Continuous unswept Rabi measurements for low amplitude signals such as those we measure are very difficult due to the sensitivity to detuning from resonance.
	A weak signal with $\sim2\pi\times\unit[20]{Hz}$ Rabi frequency is easily impacted by detunings of only a few Hertz: near-resonant Rabi measurements sense the signal \emph{total} Rabi frequency $\Omega_\text{s,R} = \sqrt{\rabis^2 + \Delta_\text{s}^2}$.
	While our calibrations are carefully set up, detuning of order $\unit[5]{Hz}$ arise from \rabic\, control errors of order $10^{-3}$.
	We estimate the signal total Rabi frequency from the measurement depicted in the spectrogram in Fig.~\ref{fig:supp_calib}. The upper and lower sidebands are split from the carrier by the static Rabi control frequency \rabic{}, which is tuned as close as we could to resonance with the signal at $\unit[10]{kHz}$. 
	The signal then drives second Rabi flopping between the dressed eigenstates, manifest as oscillation between the carrier and sidebands at the second (signal) total Rabi frequency.
	Here the spectrogram window duration was $\unit[3]{ms}$, much longer than a control Rabi cycle, but much shorter than a signal Rabi cycle, and so the dynamics induced by the signal are manifest in the spectrogram time domain, rather than as sidebands-on-sidebands. 
	These second Rabi dynamics are then retrieved from the demodulated Q channel (Fig.~\ref{fig:supp_calib}(b)), to which we have fitted a damped sinusoid from which we extract the resonant Rabi frequency $\Omega_\text{s,R}$. 
	
	We do our best to handle detuning error post-hoc, by partially solving the inverse Rabi problem~\cite{white_unpublished_nodate}.	
	Figure~\ref{fig:supp_calib}(d) shows the power spectrum of the I channel, showing it to be a pair of QAM sub-sidebands either side of a near-suppressed $\unit[10]{kHz}$ subcarrier, consistent with the modulated appearance of the primary sidebands visible in the spectrogram. 
	The relative amplitude of these upper and lower sub-sidebands  is determined by the detuning $\Delta_s$ of the test signal from resonance, i.e. the detuning from \rabic.
	While this I-channel sub-spectrum also displays $\Omega_{s,R}$ as the separation of the sub-sidebands from the subcarrier, we found it more robust to extract $\Omega_{s,R}$ from fitting the Q-channel in the time domain, as described above. 
	From these estimates of $\Delta_s$ and $\Omega_{s,R}$, we recover the \emph{resonant} Rabi frequency \rabis\, of the signal.
	Equivalently, we calibrate the amplitude of the weak test signal, using the atoms, finding a total Rabi frequency estimate of $\Omega_\text{s,R}^\text{cal}=2\pi\times\unit[22.97(1)]{Hz}$.
	Accounting for an estimated detuning of $\Delta_\text{s} = -2\pi\times\unit[2.1(6)]{Hz}$, retrieved from the relative I channel sub-sideband amplitudes in Fig.~\ref{fig:supp_calib}(d), yields the true signal Rabi frequency $\rabis^\text{cal} = \unit[22.87(6)]{Hz}$, or equivalently field amplitude $B_\text{s}^\text{cal} = \unit[3.272(9)]{nT}$.
	Our retrieved LZ amplitude of $\bs = \unit[3.34(2)]{nT}$ is within three statistical uncertainties of the calibrations, and we attribute the discrepancy to control errors, specifically to drift in the gain of the radiofrequency power amplifier in the period of several hours between the LZ measurement and the calibration shot.
	
	\section{IV: Preliminary retrievals of multi-tone spectra}
	
	While the multiparameter estimation of a single tone that we have presented is an advance towards general spectral estimation, we now show preliminary estimates of a multitone signal in a single Landau-Zener sweep. 
	
\begin{figure}
		\includegraphics[scale=0.7]{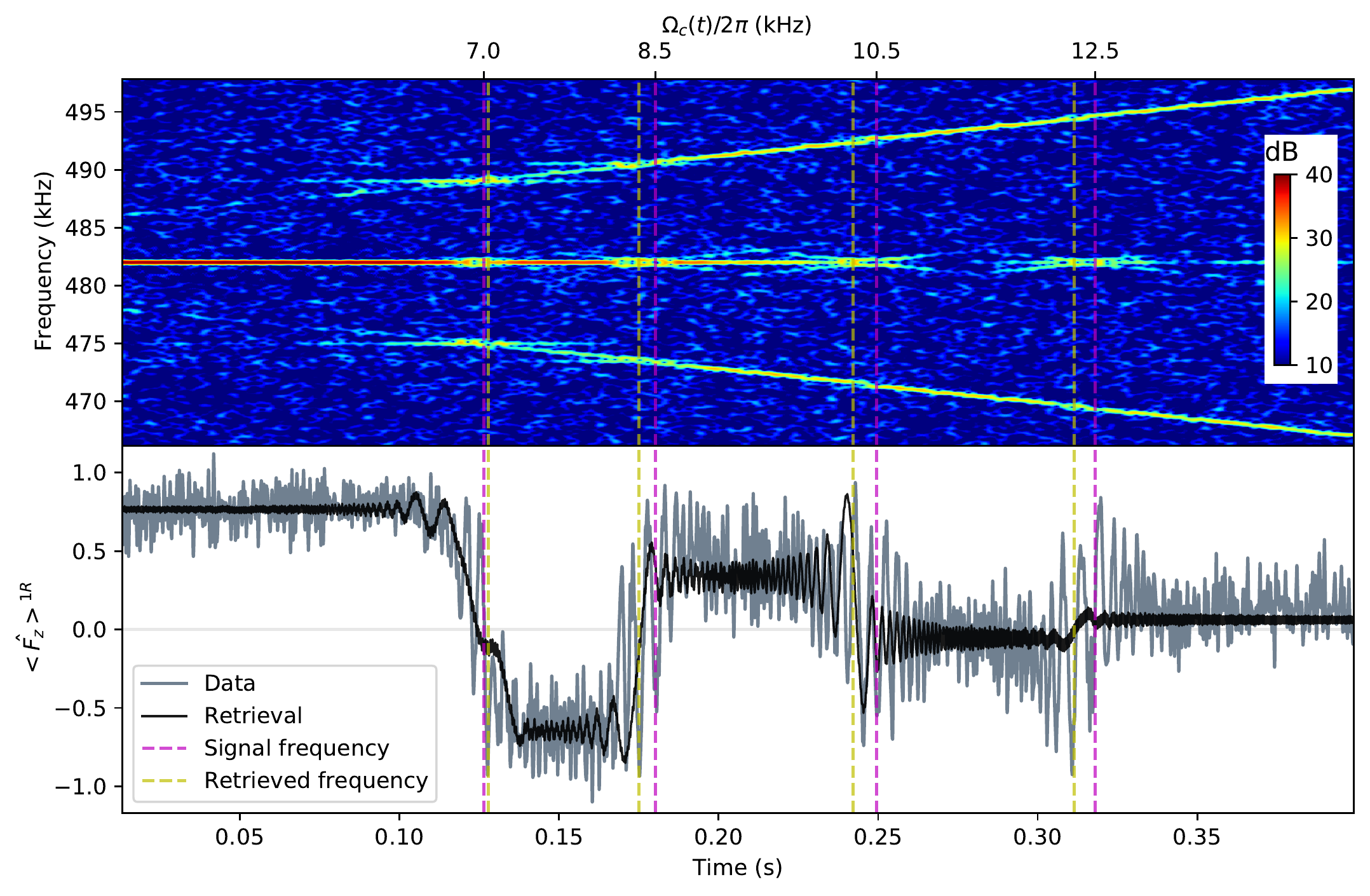}
		\caption{
			Early results for characterization of multiple signal tones at frequencies of \unit[7.0]{kHz}, \unit[8.5]{kHz}, \unit[10.5]{kHz} and \unit[12.5]{kHz} via the Landau-Zener spectrum analyzer protocol. 
			Early data is presented as both a spectrogram of power spectral density (top) with electronic noise floor at $\unit[0]{dB}$ and (bottom) frame-1R spin projection traces, with an associated inverse problem retrieval expanded to multiple signals with amplitude, signal and phase each.
			Retrieved frequencies with Rabi error correction are found to deviate by up to $\sim\unit[100]{Hz}$ with resonance times marked in purple, while expected transition times if the sweep is linear are marked in yellow.
			Discrepancies between both sets with observed avoided crossings in the spectrogram indicate considerable unhandled error in experimental implementation, with retrieval of transition structure still successful.
		}
		\label{fig:rich_spectra}
	\end{figure}

	We synthesize a chord of four tones with frequencies \unit[7.0]{kHz}, \unit[8.5]{kHz}, \unit[10.5]{kHz} and \unit[12.5]{kHz}, and equal amplitude control voltages.
	These, however, did not produce equal amplitude magnetic field components at the atoms, due to peaking in the coil driver, and parasitic reactances such as the vacuum envelope and optical breadboards, proximal to the drive coil.
	We performed a single Landau-Zener sweep of the somewhat broader band of \unit[4--15]{kHz} in a somewhat longer sweep time of $\unit[400]{ms}$, for a sweep rate of $\unit[22.5]{kHz/s}$ (c.f. \unit[20]{kHz/s} for the single-tone results), yielding the spectrogram and demodulated \Fzoner signal presented in Fig.~\ref{fig:rich_spectra}.
	Landau-Zener transitions are driven at each signal tone resonance, evident in both the spectrogram and \Fzoner\, trace.
	
	The inverse problem of retrieving four sets of signal parameters is much more challenging, not only because the increase from four parameters to thirteen, but because in this single unitary evolution the later dynamics depend on all prior resonance crossings. 	
	Nevertheless, we apply the same optimization approach to estimate parameters. 
	The frame-0 (laboratory) Hamiltonian is updated with a Fourier series for the signal,
	\begin{equation}
		\hlabt = -\hbar\freql \sigz / 2 + \hbar \rabict \cos(\freqc t) \, \sigx + \hbar \sum_{j=1}^{N} \Omega_{s,j} \cos(\omega_{s,j} t + \phi_{s,j}) \, \sigz,
	\end{equation}
	which in frame-1R retains its multiplicative character.
	We therefore can solve the forward problem using the same Bloch equation solver but now with its time-dependent Rabi vector containing the Fourier sum viz $\vec{\Omega}^\text{1R} = 2\sum_{j=1}^n \Omega_\text{s,j} \cos(\omega_\text{s,j} t + \phi_\text{s,j})\hat{x} + \rabict\hat{z}$. 
	The global optimizer then optimizes for parameters $\omega_{s,j}$, $\Omega_{s,j}$, $\phi_{s,j}$ and the initial state \Fzoneri. 
	The solution presented in Fig.~\ref{fig:rich_spectra} was obtained after $7.5\times10^4$ steps, taking $\unit[108.5]{mins}$ instead of the $\unit[<10]{mins}$ required for single-tone retrievals.
	The optimizer successfully detected $N=2$ tone signals with the same lax constraints as the single-tone results (signal frequencies within sweep, and amplitude within broad positive range as above). 
	On $N=4$ tones, these lax constraints typically resulted in the optimizer returning two degenerate frequencies with sharing of amplitude, and consequently missing the weakest transition. 
	However, once constrained to broad, contiguous bounds for $\omega_{s,j}$, the optimizer always converged to the vicinity of the true signal frequencies, even when we assigned unequal amplitudes to the tones. 
	We emphasize that the optimizer always operated on the entire timeseries, and the only change was to add parameter constraints for $N=4$.
	While we set these constraints manually, it is clear from  Fig.~\ref{fig:rich_spectra} that a simple heuristic, for example based of fitting a piecewise-constant function to \Fzoner, would suffice to automatically determine such constraints, at least for sparse spectra comprising well-separated tones.
	
	Retrieved frequencies of $\omega_{s,1}=2\pi\times\unit[7.11350(16)]{kHz}$, $\omega_{s,2}=2\pi\times\unit[8.44775(16)]{kHz}$, $\omega_{s,3}=2\pi\times\unit[10.33979(16)]{kHz}$ and $\omega_{s,4}=2\pi\times\unit[12.40672(16)]{kHz}$ for the four tones respectively show disagreement of up to $\unit[160]{Hz}$ with commanded signal frequencies.
	The identical commanded amplitudes here were much stronger than the genuinely weak signals used in our primary single-tone result, and drive near-adiabatic transitions, visible as the near complete inversion of the initial state at the first resonance crossing, and as characteristic oscillation \emph{during} the resonance crossing thereafter. 
	
	Retrieved amplitudes were not as encouraging as the frequencies, falling between $\Omega_{s,2}=2\pi\times\unit[91]{Hz}$ and $\Omega_{s,4}=2\pi\times\unit[16]{Hz}$.
	For comparison, here the adiabaticity threshold is $\sqrt{\lambda}=2\pi\times\unit[59.8]{Hz}$.
	We posit that this inconsistency in amplitude arises from the even more oscillatory cost landscape for near-adiabatic transitions. 
	Even though in principle we expect the optimizer to be able to retrieve parameters in this regime, the combined burden of dimensionality and difficulty of finding the global minimum in this landscape leads to reasonably consistent estimates of signal frequencies but not amplitudes. 
	Instead, the optimizer frequently settles on weaker signal amplitudes, apparent as overly smooth trajectories through most, but not all, of the resonances.
	Further, we suspect that the covariance between frequency and amplitude estimates contributes to the frequency estimate residuals. 
	In these multi-tone data sets, we had not yet linearized the Rabi sweep, however we were able to take advantage of the self-calibrating property of Rabi error correction, and correct most, but not all, of the time-series for Rabi error. 
	
	These caveats highlight why we consider these measurements preliminary.
	We are confident, however, that this optimization-based approach to quantum spectrum analysis is viable for sparse spectra, and we are endeavoring to extend the precise and sensitive retrievals demonstrated for single-tone signals to retrieve multi-tone spectra in the metrologically-significant limit of low amplitudes.


\begin{thebibliography}{57}%
		\makeatletter
		\providecommand \@ifxundefined [1]{%
			\@ifx{#1\undefined}
		}%
		\providecommand \@ifnum [1]{%
			\ifnum #1\expandafter \@firstoftwo
			\else \expandafter \@secondoftwo
			\fi
		}%
		\providecommand \@ifx [1]{%
			\ifx #1\expandafter \@firstoftwo
			\else \expandafter \@secondoftwo
			\fi
		}%
		\providecommand \natexlab [1]{#1}%
		\providecommand \enquote  [1]{``#1''}%
		\providecommand \bibnamefont  [1]{#1}%
		\providecommand \bibfnamefont [1]{#1}%
		\providecommand \citenamefont [1]{#1}%
		\providecommand \href@noop [0]{\@secondoftwo}%
		\providecommand \href [0]{\begingroup \@sanitize@url \@href}%
		\providecommand \@href[1]{\@@startlink{#1}\@@href}%
		\providecommand \@@href[1]{\endgroup#1\@@endlink}%
		\providecommand \@sanitize@url [0]{\catcode `\\12\catcode `\$12\catcode
			`\&12\catcode `\#12\catcode `\^12\catcode `\_12\catcode `\%12\relax}%
		\providecommand \@@startlink[1]{}%
		\providecommand \@@endlink[0]{}%
		\providecommand \url  [0]{\begingroup\@sanitize@url \@url }%
		\providecommand \@url [1]{\endgroup\@href {#1}{\urlprefix }}%
		\providecommand \urlprefix  [0]{URL }%
		\providecommand \Eprint [0]{\href }%
		\providecommand \doibase [0]{https://doi.org/}%
		\providecommand \selectlanguage [0]{\@gobble}%
		\providecommand \bibinfo  [0]{\@secondoftwo}%
		\providecommand \bibfield  [0]{\@secondoftwo}%
		\providecommand \translation [1]{[#1]}%
		\providecommand \BibitemOpen [0]{}%
		\providecommand \bibitemStop [0]{}%
		\providecommand \bibitemNoStop [0]{.\EOS\space}%
		\providecommand \EOS [0]{\spacefactor3000\relax}%
		\providecommand \BibitemShut  [1]{\csname bibitem#1\endcsname}%
		\let\auto@bib@innerbib\@empty
		\bibitem [{\citenamefont {Degen}\ \emph {et~al.}(2017)\citenamefont {Degen},
			\citenamefont {Reinhard},\ and\ \citenamefont
			{Cappellaro}}]{degen_quantum_2017}%
		\BibitemOpen
		\bibfield  {author} {\bibinfo {author} {\bibfnamefont {C.~L.}\ \bibnamefont
				{Degen}}, \bibinfo {author} {\bibfnamefont {F.}~\bibnamefont {Reinhard}},\
			and\ \bibinfo {author} {\bibfnamefont {P.}~\bibnamefont {Cappellaro}},\
		}\bibfield  {title} {\bibinfo {title} {Quantum sensing},\ }\href
		{https://doi.org/10.1103/RevModPhys.89.035002} {\bibfield  {journal}
			{\bibinfo  {journal} {Rev. Mod. Phys.}\ }\textbf {\bibinfo {volume} {89}},\
			\bibinfo {pages} {035002} (\bibinfo {year} {2017})}\BibitemShut {NoStop}%
		\bibitem [{\citenamefont {Engelson}(1984)}]{engelson_modern_1984}%
		\BibitemOpen
		\bibfield  {author} {\bibinfo {author} {\bibfnamefont {M.}~\bibnamefont
				{Engelson}},\ }\href@noop {} {\emph {\bibinfo {title} {Modern {{Spectrum
							Analyzer Theory}} and {{Applications}}}}}\ (\bibinfo  {publisher} {{Artech
				House}},\ \bibinfo {year} {1984})\BibitemShut {NoStop}%
		\bibitem [{\citenamefont {Koenig}\ \emph {et~al.}(1946)\citenamefont {Koenig},
			\citenamefont {Dunn},\ and\ \citenamefont {Lacy}}]{koenig_sound_1946}%
		\BibitemOpen
		\bibfield  {author} {\bibinfo {author} {\bibfnamefont {W.}~\bibnamefont
				{Koenig}}, \bibinfo {author} {\bibfnamefont {H.~K.}\ \bibnamefont {Dunn}},\
			and\ \bibinfo {author} {\bibfnamefont {L.~Y.}\ \bibnamefont {Lacy}},\
		}\bibfield  {title} {\bibinfo {title} {The {{Sound Spectrograph}}},\ }\href
		{https://doi.org/10.1121/1.1916342} {\bibfield  {journal} {\bibinfo
				{journal} {J. Acoust. Soc. Am.}\ }\textbf {\bibinfo {volume} {18}},\ \bibinfo
			{pages} {19} (\bibinfo {year} {1946})}\BibitemShut {NoStop}%
		\bibitem [{\citenamefont {Thomson}(1878)}]{thomson_iv_1878}%
		\BibitemOpen
		\bibfield  {author} {\bibinfo {author} {\bibfnamefont {W.}~\bibnamefont
				{Thomson}},\ }\bibfield  {title} {\bibinfo {title} {{{IV}}. {{Harmonic}}
				analyzer},\ }\href {https://doi.org/10.1098/rspl.1878.0062} {\bibfield
			{journal} {\bibinfo  {journal} {Proc. R. Soc}\ }\textbf {\bibinfo {volume}
				{27}},\ \bibinfo {pages} {371} (\bibinfo {year} {1878})}\BibitemShut
		{NoStop}%
		\bibitem [{\citenamefont {Michelson}\ and\ \citenamefont
			{Stratton}(1898)}]{michelson_new_1898}%
		\BibitemOpen
		\bibfield  {author} {\bibinfo {author} {\bibfnamefont {A.~A.}\ \bibnamefont
				{Michelson}}\ and\ \bibinfo {author} {\bibfnamefont {S.~W.}\ \bibnamefont
				{Stratton}},\ }\bibfield  {title} {\bibinfo {title} {A new harmonic
				analyzer},\ }\href {https://doi.org/10.2475/ajs.s4-5.25.1} {\bibfield
			{journal} {\bibinfo  {journal} {Am. J. Sci.}\ }\textbf {\bibinfo {volume}
				{s4-5}},\ \bibinfo {pages} {1} (\bibinfo {year} {1898})}\BibitemShut
		{NoStop}%
		\bibitem [{\citenamefont {Kotler}\ \emph {et~al.}(2013)\citenamefont {Kotler},
			\citenamefont {Akerman}, \citenamefont {Glickman},\ and\ \citenamefont
			{Ozeri}}]{kotler_nonlinear_2013}%
		\BibitemOpen
		\bibfield  {author} {\bibinfo {author} {\bibfnamefont {S.}~\bibnamefont
				{Kotler}}, \bibinfo {author} {\bibfnamefont {N.}~\bibnamefont {Akerman}},
			\bibinfo {author} {\bibfnamefont {Y.}~\bibnamefont {Glickman}},\ and\
			\bibinfo {author} {\bibfnamefont {R.}~\bibnamefont {Ozeri}},\ }\bibfield
		{title} {\bibinfo {title} {Nonlinear {{Single-Spin Spectrum Analyzer}}},\
		}\href {https://doi.org/10.1103/PhysRevLett.110.110503} {\bibfield  {journal}
			{\bibinfo  {journal} {Phys. Rev. Lett.}\ }\textbf {\bibinfo {volume} {110}},\
			\bibinfo {pages} {110503} (\bibinfo {year} {2013})}\BibitemShut {NoStop}%
		\bibitem [{\citenamefont {Chipaux}\ \emph {et~al.}(2015)\citenamefont
			{Chipaux}, \citenamefont {Toraille}, \citenamefont {Larat}, \citenamefont
			{Morvan}, \citenamefont {Pezzagna}, \citenamefont {Meijer},\ and\
			\citenamefont {Debuisschert}}]{chipaux_wide_2015}%
		\BibitemOpen
		\bibfield  {author} {\bibinfo {author} {\bibfnamefont {M.}~\bibnamefont
				{Chipaux}}, \bibinfo {author} {\bibfnamefont {L.}~\bibnamefont {Toraille}},
			\bibinfo {author} {\bibfnamefont {C.}~\bibnamefont {Larat}}, \bibinfo
			{author} {\bibfnamefont {L.}~\bibnamefont {Morvan}}, \bibinfo {author}
			{\bibfnamefont {S.}~\bibnamefont {Pezzagna}}, \bibinfo {author}
			{\bibfnamefont {J.}~\bibnamefont {Meijer}},\ and\ \bibinfo {author}
			{\bibfnamefont {T.}~\bibnamefont {Debuisschert}},\ }\bibfield  {title}
		{\bibinfo {title} {Wide bandwidth instantaneous radio frequency spectrum
				analyzer based on nitrogen vacancy centers in diamond},\ }\href
		{https://doi.org/10.1063/1.4936758} {\bibfield  {journal} {\bibinfo
				{journal} {Appl. Phys. Lett.}\ }\textbf {\bibinfo {volume} {107}},\ \bibinfo
			{pages} {233502} (\bibinfo {year} {2015})}\BibitemShut {NoStop}%
		\bibitem [{\citenamefont {Magaletti}\ \emph {et~al.}(2022)\citenamefont
			{Magaletti}, \citenamefont {Mayer}, \citenamefont {Roch},\ and\ \citenamefont
			{Debuisschert}}]{magaletti_quantum_2022}%
		\BibitemOpen
		\bibfield  {author} {\bibinfo {author} {\bibfnamefont {S.}~\bibnamefont
				{Magaletti}}, \bibinfo {author} {\bibfnamefont {L.}~\bibnamefont {Mayer}},
			\bibinfo {author} {\bibfnamefont {J.-F.}\ \bibnamefont {Roch}},\ and\
			\bibinfo {author} {\bibfnamefont {T.}~\bibnamefont {Debuisschert}},\
		}\bibfield  {title} {\bibinfo {title} {A quantum radio frequency signal
				analyzer based on nitrogen vacancy centers in diamond},\ }\href
		{https://doi.org/10.1038/s44172-022-00017-4} {\bibfield  {journal} {\bibinfo
				{journal} {Commun Eng}\ }\textbf {\bibinfo {volume} {1}},\ \bibinfo {pages}
			{1} (\bibinfo {year} {2022})}\BibitemShut {NoStop}%
		\bibitem [{\citenamefont {Haeberlen}\ and\ \citenamefont
			{Waugh}(1968)}]{haeberlen_coherent_1968}%
		\BibitemOpen
		\bibfield  {author} {\bibinfo {author} {\bibfnamefont {U.}~\bibnamefont
				{Haeberlen}}\ and\ \bibinfo {author} {\bibfnamefont {J.~S.}\ \bibnamefont
				{Waugh}},\ }\bibfield  {title} {\bibinfo {title} {Coherent {{Averaging
						Effects}} in {{Magnetic Resonance}}},\ }\href
		{https://doi.org/10.1103/PhysRev.175.453} {\bibfield  {journal} {\bibinfo
				{journal} {Phys. Rev.}\ }\textbf {\bibinfo {volume} {175}},\ \bibinfo {pages}
			{453} (\bibinfo {year} {1968})}\BibitemShut {NoStop}%
		\bibitem [{\citenamefont {Facchi}\ \emph {et~al.}(2004)\citenamefont {Facchi},
			\citenamefont {Lidar},\ and\ \citenamefont
			{Pascazio}}]{facchi_unification_2004}%
		\BibitemOpen
		\bibfield  {author} {\bibinfo {author} {\bibfnamefont {P.}~\bibnamefont
				{Facchi}}, \bibinfo {author} {\bibfnamefont {D.~A.}\ \bibnamefont {Lidar}},\
			and\ \bibinfo {author} {\bibfnamefont {S.}~\bibnamefont {Pascazio}},\
		}\bibfield  {title} {\bibinfo {title} {Unification of dynamical decoupling
				and the quantum {{Zeno}} effect},\ }\href
		{https://doi.org/10.1103/PhysRevA.69.032314} {\bibfield  {journal} {\bibinfo
				{journal} {Phys. Rev. A}\ }\textbf {\bibinfo {volume} {69}},\ \bibinfo
			{pages} {032314} (\bibinfo {year} {2004})}\BibitemShut {NoStop}%
		\bibitem [{\citenamefont {Cai}\ \emph {et~al.}(2012)\citenamefont {Cai},
			\citenamefont {Naydenov}, \citenamefont {Pfeiffer}, \citenamefont
			{McGuinness}, \citenamefont {Jahnke}, \citenamefont {Jelezko}, \citenamefont
			{Plenio},\ and\ \citenamefont {Retzker}}]{cai_robust_2012}%
		\BibitemOpen
		\bibfield  {author} {\bibinfo {author} {\bibfnamefont {J.-M.}\ \bibnamefont
				{Cai}}, \bibinfo {author} {\bibfnamefont {B.}~\bibnamefont {Naydenov}},
			\bibinfo {author} {\bibfnamefont {R.}~\bibnamefont {Pfeiffer}}, \bibinfo
			{author} {\bibfnamefont {L.~P.}\ \bibnamefont {McGuinness}}, \bibinfo
			{author} {\bibfnamefont {K.~D.}\ \bibnamefont {Jahnke}}, \bibinfo {author}
			{\bibfnamefont {F.}~\bibnamefont {Jelezko}}, \bibinfo {author} {\bibfnamefont
				{M.~B.}\ \bibnamefont {Plenio}},\ and\ \bibinfo {author} {\bibfnamefont
				{A.}~\bibnamefont {Retzker}},\ }\bibfield  {title} {\bibinfo {title} {Robust
				dynamical decoupling with concatenated continuous driving},\ }\href
		{https://doi.org/10.1088/1367-2630/14/11/113023} {\bibfield  {journal}
			{\bibinfo  {journal} {New J. Phys.}\ }\textbf {\bibinfo {volume} {14}},\
			\bibinfo {pages} {113023} (\bibinfo {year} {2012})}\BibitemShut {NoStop}%
		\bibitem [{\citenamefont {Ajoy}\ \emph {et~al.}(2017)\citenamefont {Ajoy},
			\citenamefont {Liu}, \citenamefont {Saha}, \citenamefont {Marseglia},
			\citenamefont {Jaskula}, \citenamefont {Bissbort},\ and\ \citenamefont
			{Cappellaro}}]{ajoy_quantum_2017}%
		\BibitemOpen
		\bibfield  {author} {\bibinfo {author} {\bibfnamefont {A.}~\bibnamefont
				{Ajoy}}, \bibinfo {author} {\bibfnamefont {Y.-X.}\ \bibnamefont {Liu}},
			\bibinfo {author} {\bibfnamefont {K.}~\bibnamefont {Saha}}, \bibinfo {author}
			{\bibfnamefont {L.}~\bibnamefont {Marseglia}}, \bibinfo {author}
			{\bibfnamefont {J.-C.}\ \bibnamefont {Jaskula}}, \bibinfo {author}
			{\bibfnamefont {U.}~\bibnamefont {Bissbort}},\ and\ \bibinfo {author}
			{\bibfnamefont {P.}~\bibnamefont {Cappellaro}},\ }\bibfield  {title}
		{\bibinfo {title} {Quantum interpolation for high-resolution sensing},\
		}\href {https://doi.org/10.1073/pnas.1610835114} {\bibfield  {journal}
			{\bibinfo  {journal} {Proc. Natl. Acad. Sci. U.S.A.}\ }\textbf {\bibinfo
				{volume} {114}},\ \bibinfo {pages} {2149} (\bibinfo {year}
			{2017})}\BibitemShut {NoStop}%
		\bibitem [{\citenamefont {Schmitt}\ \emph {et~al.}(2017)\citenamefont
			{Schmitt}, \citenamefont {Gefen}, \citenamefont {St{\"u}rner}, \citenamefont
			{Unden}, \citenamefont {Wolff}, \citenamefont {M{\"u}ller}, \citenamefont
			{Scheuer}, \citenamefont {Naydenov}, \citenamefont {Markham}, \citenamefont
			{Pezzagna}, \citenamefont {Meijer}, \citenamefont {Schwarz}, \citenamefont
			{Plenio}, \citenamefont {Retzker}, \citenamefont {McGuinness},\ and\
			\citenamefont {Jelezko}}]{schmitt_submillihertz_2017}%
		\BibitemOpen
		\bibfield  {author} {\bibinfo {author} {\bibfnamefont {S.}~\bibnamefont
				{Schmitt}}, \bibinfo {author} {\bibfnamefont {T.}~\bibnamefont {Gefen}},
			\bibinfo {author} {\bibfnamefont {F.~M.}\ \bibnamefont {St{\"u}rner}},
			\bibinfo {author} {\bibfnamefont {T.}~\bibnamefont {Unden}}, \bibinfo
			{author} {\bibfnamefont {G.}~\bibnamefont {Wolff}}, \bibinfo {author}
			{\bibfnamefont {C.}~\bibnamefont {M{\"u}ller}}, \bibinfo {author}
			{\bibfnamefont {J.}~\bibnamefont {Scheuer}}, \bibinfo {author} {\bibfnamefont
				{B.}~\bibnamefont {Naydenov}}, \bibinfo {author} {\bibfnamefont
				{M.}~\bibnamefont {Markham}}, \bibinfo {author} {\bibfnamefont
				{S.}~\bibnamefont {Pezzagna}}, \bibinfo {author} {\bibfnamefont
				{J.}~\bibnamefont {Meijer}}, \bibinfo {author} {\bibfnamefont
				{I.}~\bibnamefont {Schwarz}}, \bibinfo {author} {\bibfnamefont
				{M.}~\bibnamefont {Plenio}}, \bibinfo {author} {\bibfnamefont
				{A.}~\bibnamefont {Retzker}}, \bibinfo {author} {\bibfnamefont {L.~P.}\
				\bibnamefont {McGuinness}},\ and\ \bibinfo {author} {\bibfnamefont
				{F.}~\bibnamefont {Jelezko}},\ }\bibfield  {title} {\bibinfo {title}
			{Submillihertz magnetic spectroscopy performed with a nanoscale quantum
				sensor},\ }\href {https://doi.org/10.1126/science.aam5532} {\bibfield
			{journal} {\bibinfo  {journal} {Science}\ }\textbf {\bibinfo {volume}
				{356}},\ \bibinfo {pages} {832} (\bibinfo {year} {2017})}\BibitemShut
		{NoStop}%
		\bibitem [{\citenamefont {Boss}\ \emph {et~al.}(2017)\citenamefont {Boss},
			\citenamefont {Cujia}, \citenamefont {Zopes},\ and\ \citenamefont
			{Degen}}]{boss_quantum_2017-1}%
		\BibitemOpen
		\bibfield  {author} {\bibinfo {author} {\bibfnamefont {J.~M.}\ \bibnamefont
				{Boss}}, \bibinfo {author} {\bibfnamefont {K.~S.}\ \bibnamefont {Cujia}},
			\bibinfo {author} {\bibfnamefont {J.}~\bibnamefont {Zopes}},\ and\ \bibinfo
			{author} {\bibfnamefont {C.~L.}\ \bibnamefont {Degen}},\ }\bibfield  {title}
		{\bibinfo {title} {Quantum sensing with arbitrary frequency resolution},\
		}\href {https://doi.org/10.1126/science.aam7009} {\bibfield  {journal}
			{\bibinfo  {journal} {Science}\ }\textbf {\bibinfo {volume} {356}},\ \bibinfo
			{pages} {837} (\bibinfo {year} {2017})}\BibitemShut {NoStop}%
		\bibitem [{\citenamefont {Trypogeorgos}\ \emph {et~al.}(2018)\citenamefont
			{Trypogeorgos}, \citenamefont {{Vald{\'e}s-Curiel}}, \citenamefont
			{Lundblad},\ and\ \citenamefont {Spielman}}]{trypogeorgos_synthetic_2018}%
		\BibitemOpen
		\bibfield  {author} {\bibinfo {author} {\bibfnamefont {D.}~\bibnamefont
				{Trypogeorgos}}, \bibinfo {author} {\bibfnamefont {A.}~\bibnamefont
				{{Vald{\'e}s-Curiel}}}, \bibinfo {author} {\bibfnamefont {N.}~\bibnamefont
				{Lundblad}},\ and\ \bibinfo {author} {\bibfnamefont {I.~B.}\ \bibnamefont
				{Spielman}},\ }\bibfield  {title} {\bibinfo {title} {Synthetic clock
				transitions via continuous dynamical decoupling},\ }\href
		{https://doi.org/10.1103/PhysRevA.97.013407} {\bibfield  {journal} {\bibinfo
				{journal} {Phys. Rev. A}\ }\textbf {\bibinfo {volume} {97}},\ \bibinfo
			{pages} {013407} (\bibinfo {year} {2018})}\BibitemShut {NoStop}%
		\bibitem [{\citenamefont {Anderson}\ \emph {et~al.}(2018)\citenamefont
			{Anderson}, \citenamefont {Kewming},\ and\ \citenamefont
			{Turner}}]{anderson_continuously_2018}%
		\BibitemOpen
		\bibfield  {author} {\bibinfo {author} {\bibfnamefont {R.~P.}\ \bibnamefont
				{Anderson}}, \bibinfo {author} {\bibfnamefont {M.~J.}\ \bibnamefont
				{Kewming}},\ and\ \bibinfo {author} {\bibfnamefont {L.~D.}\ \bibnamefont
				{Turner}},\ }\bibfield  {title} {\bibinfo {title} {Continuously observing a
				dynamically decoupled spin-1 quantum gas},\ }\href
		{https://doi.org/10.1103/PhysRevA.97.013408} {\bibfield  {journal} {\bibinfo
				{journal} {Phys. Rev. A}\ }\textbf {\bibinfo {volume} {97}},\ \bibinfo
			{pages} {013408} (\bibinfo {year} {2018})}\BibitemShut {NoStop}%
		\bibitem [{\citenamefont {Staudenmaier}\ \emph {et~al.}(2021)\citenamefont
			{Staudenmaier}, \citenamefont {Schmitt}, \citenamefont {McGuinness},\ and\
			\citenamefont {Jelezko}}]{staudenmaier_phase-sensitive_2021}%
		\BibitemOpen
		\bibfield  {author} {\bibinfo {author} {\bibfnamefont {N.}~\bibnamefont
				{Staudenmaier}}, \bibinfo {author} {\bibfnamefont {S.}~\bibnamefont
				{Schmitt}}, \bibinfo {author} {\bibfnamefont {L.~P.}\ \bibnamefont
				{McGuinness}},\ and\ \bibinfo {author} {\bibfnamefont {F.}~\bibnamefont
				{Jelezko}},\ }\bibfield  {title} {\bibinfo {title} {Phase-sensitive quantum
				spectroscopy with high-frequency resolution},\ }\href
		{https://doi.org/10.1103/PhysRevA.104.L020602} {\bibfield  {journal}
			{\bibinfo  {journal} {Phys. Rev. A}\ }\textbf {\bibinfo {volume} {104}},\
			\bibinfo {pages} {L020602} (\bibinfo {year} {2021})}\BibitemShut {NoStop}%
		\bibitem [{\citenamefont {Meinel}\ \emph {et~al.}(2021)\citenamefont {Meinel},
			\citenamefont {Vorobyov}, \citenamefont {Yavkin}, \citenamefont {Dasari},
			\citenamefont {Sumiya}, \citenamefont {Onoda}, \citenamefont {Isoya},\ and\
			\citenamefont {Wrachtrup}}]{meinel_heterodyne_2021}%
		\BibitemOpen
		\bibfield  {author} {\bibinfo {author} {\bibfnamefont {J.}~\bibnamefont
				{Meinel}}, \bibinfo {author} {\bibfnamefont {V.}~\bibnamefont {Vorobyov}},
			\bibinfo {author} {\bibfnamefont {B.}~\bibnamefont {Yavkin}}, \bibinfo
			{author} {\bibfnamefont {D.}~\bibnamefont {Dasari}}, \bibinfo {author}
			{\bibfnamefont {H.}~\bibnamefont {Sumiya}}, \bibinfo {author} {\bibfnamefont
				{S.}~\bibnamefont {Onoda}}, \bibinfo {author} {\bibfnamefont
				{J.}~\bibnamefont {Isoya}},\ and\ \bibinfo {author} {\bibfnamefont
				{J.}~\bibnamefont {Wrachtrup}},\ }\bibfield  {title} {\bibinfo {title}
			{Heterodyne sensing of microwaves with a quantum sensor},\ }\href
		{https://doi.org/10.1038/s41467-021-22714-y} {\bibfield  {journal} {\bibinfo
				{journal} {Nat Commun}\ }\textbf {\bibinfo {volume} {12}},\ \bibinfo {pages}
			{2737} (\bibinfo {year} {2021})}\BibitemShut {NoStop}%
		\bibitem [{\citenamefont {Khodjasteh}\ and\ \citenamefont
			{Viola}(2009)}]{khodjasteh_dynamically_2009}%
		\BibitemOpen
		\bibfield  {author} {\bibinfo {author} {\bibfnamefont {K.}~\bibnamefont
				{Khodjasteh}}\ and\ \bibinfo {author} {\bibfnamefont {L.}~\bibnamefont
				{Viola}},\ }\bibfield  {title} {\bibinfo {title} {Dynamically
				{{Error-Corrected Gates}} for {{Universal Quantum Computation}}},\ }\href
		{https://doi.org/10.1103/PhysRevLett.102.080501} {\bibfield  {journal}
			{\bibinfo  {journal} {Phys. Rev. Lett.}\ }\textbf {\bibinfo {volume} {102}},\
			\bibinfo {pages} {080501} (\bibinfo {year} {2009})}\BibitemShut {NoStop}%
		\bibitem [{\citenamefont {Ban}(1998)}]{ban_photon-echo_1998}%
		\BibitemOpen
		\bibfield  {author} {\bibinfo {author} {\bibfnamefont {M.}~\bibnamefont
				{Ban}},\ }\bibfield  {title} {\bibinfo {title} {Photon-echo technique for
				reducing the decoherence of a quantum bit},\ }\href
		{https://doi.org/10.1080/09500349808231241} {\bibfield  {journal} {\bibinfo
				{journal} {J. Mod. Opt.}\ }\textbf {\bibinfo {volume} {45}},\ \bibinfo
			{pages} {2315} (\bibinfo {year} {1998})}\BibitemShut {NoStop}%
		\bibitem [{\citenamefont {Taylor}\ \emph {et~al.}(2008)\citenamefont {Taylor},
			\citenamefont {Cappellaro}, \citenamefont {Childress}, \citenamefont {Jiang},
			\citenamefont {Budker}, \citenamefont {Hemmer}, \citenamefont {Yacoby},
			\citenamefont {Walsworth},\ and\ \citenamefont
			{Lukin}}]{taylor_high-sensitivity_2008}%
		\BibitemOpen
		\bibfield  {author} {\bibinfo {author} {\bibfnamefont {J.~M.}\ \bibnamefont
				{Taylor}}, \bibinfo {author} {\bibfnamefont {P.}~\bibnamefont {Cappellaro}},
			\bibinfo {author} {\bibfnamefont {L.}~\bibnamefont {Childress}}, \bibinfo
			{author} {\bibfnamefont {L.}~\bibnamefont {Jiang}}, \bibinfo {author}
			{\bibfnamefont {D.}~\bibnamefont {Budker}}, \bibinfo {author} {\bibfnamefont
				{P.~R.}\ \bibnamefont {Hemmer}}, \bibinfo {author} {\bibfnamefont
				{A.}~\bibnamefont {Yacoby}}, \bibinfo {author} {\bibfnamefont
				{R.}~\bibnamefont {Walsworth}},\ and\ \bibinfo {author} {\bibfnamefont
				{M.~D.}\ \bibnamefont {Lukin}},\ }\bibfield  {title} {\bibinfo {title}
			{High-sensitivity diamond magnetometer with nanoscale resolution},\ }\href
		{https://doi.org/10.1038/nphys1075} {\bibfield  {journal} {\bibinfo
				{journal} {Nature Phys}\ }\textbf {\bibinfo {volume} {4}},\ \bibinfo {pages}
			{810} (\bibinfo {year} {2008})}\BibitemShut {NoStop}%
		\bibitem [{\citenamefont {Kotler}\ \emph {et~al.}(2011)\citenamefont {Kotler},
			\citenamefont {Akerman}, \citenamefont {Glickman}, \citenamefont {Keselman},\
			and\ \citenamefont {Ozeri}}]{kotler_single-ion_2011}%
		\BibitemOpen
		\bibfield  {author} {\bibinfo {author} {\bibfnamefont {S.}~\bibnamefont
				{Kotler}}, \bibinfo {author} {\bibfnamefont {N.}~\bibnamefont {Akerman}},
			\bibinfo {author} {\bibfnamefont {Y.}~\bibnamefont {Glickman}}, \bibinfo
			{author} {\bibfnamefont {A.}~\bibnamefont {Keselman}},\ and\ \bibinfo
			{author} {\bibfnamefont {R.}~\bibnamefont {Ozeri}},\ }\bibfield  {title}
		{\bibinfo {title} {Single-ion quantum lock-in amplifier},\ }\href
		{https://doi.org/10.1038/nature10010} {\bibfield  {journal} {\bibinfo
				{journal} {Nature}\ }\textbf {\bibinfo {volume} {473}},\ \bibinfo {pages}
			{61} (\bibinfo {year} {2011})}\BibitemShut {NoStop}%
		\bibitem [{\citenamefont {Frey}\ \emph {et~al.}(2017)\citenamefont {Frey},
			\citenamefont {Mavadia}, \citenamefont {Norris}, \citenamefont {{de
					Ferranti}}, \citenamefont {Lucarelli}, \citenamefont {Viola},\ and\
			\citenamefont {Biercuk}}]{frey_application_2017}%
		\BibitemOpen
		\bibfield  {author} {\bibinfo {author} {\bibfnamefont {V.~M.}\ \bibnamefont
				{Frey}}, \bibinfo {author} {\bibfnamefont {S.}~\bibnamefont {Mavadia}},
			\bibinfo {author} {\bibfnamefont {L.~M.}\ \bibnamefont {Norris}}, \bibinfo
			{author} {\bibfnamefont {W.}~\bibnamefont {{de Ferranti}}}, \bibinfo {author}
			{\bibfnamefont {D.}~\bibnamefont {Lucarelli}}, \bibinfo {author}
			{\bibfnamefont {L.}~\bibnamefont {Viola}},\ and\ \bibinfo {author}
			{\bibfnamefont {M.~J.}\ \bibnamefont {Biercuk}},\ }\bibfield  {title}
		{\bibinfo {title} {Application of optimal band-limited control protocols to
				quantum noise sensing},\ }\href {https://doi.org/10.1038/s41467-017-02298-2}
		{\bibfield  {journal} {\bibinfo  {journal} {Nat Commun}\ }\textbf {\bibinfo
				{volume} {8}},\ \bibinfo {pages} {2189} (\bibinfo {year} {2017})}\BibitemShut
		{NoStop}%
		\bibitem [{\citenamefont {Zhang}\ \emph {et~al.}(2017)\citenamefont {Zhang},
			\citenamefont {Arai}, \citenamefont {Belthangady}, \citenamefont {Jaskula},\
			and\ \citenamefont {Walsworth}}]{zhang_selective_2017}%
		\BibitemOpen
		\bibfield  {author} {\bibinfo {author} {\bibfnamefont {H.}~\bibnamefont
				{Zhang}}, \bibinfo {author} {\bibfnamefont {K.}~\bibnamefont {Arai}},
			\bibinfo {author} {\bibfnamefont {C.}~\bibnamefont {Belthangady}}, \bibinfo
			{author} {\bibfnamefont {J.-C.}\ \bibnamefont {Jaskula}},\ and\ \bibinfo
			{author} {\bibfnamefont {R.~L.}\ \bibnamefont {Walsworth}},\ }\bibfield
		{title} {\bibinfo {title} {Selective addressing of solid-state spins at the
				nanoscale via magnetic resonance frequency encoding},\ }\href
		{https://doi.org/10.1038/s41534-017-0033-3} {\bibfield  {journal} {\bibinfo
				{journal} {npj Quantum Inf}\ }\textbf {\bibinfo {volume} {3}},\ \bibinfo
			{pages} {1} (\bibinfo {year} {2017})}\BibitemShut {NoStop}%
		\bibitem [{\citenamefont {Jasperse}\ \emph {et~al.}(2017)\citenamefont
			{Jasperse}, \citenamefont {Kewming}, \citenamefont {Fischer}, \citenamefont
			{Pakkiam}, \citenamefont {Anderson},\ and\ \citenamefont
			{Turner}}]{jasperse_continuous_2017}%
		\BibitemOpen
		\bibfield  {author} {\bibinfo {author} {\bibfnamefont {M.}~\bibnamefont
				{Jasperse}}, \bibinfo {author} {\bibfnamefont {M.~J.}\ \bibnamefont
				{Kewming}}, \bibinfo {author} {\bibfnamefont {S.~N.}\ \bibnamefont
				{Fischer}}, \bibinfo {author} {\bibfnamefont {P.}~\bibnamefont {Pakkiam}},
			\bibinfo {author} {\bibfnamefont {R.~P.}\ \bibnamefont {Anderson}},\ and\
			\bibinfo {author} {\bibfnamefont {L.~D.}\ \bibnamefont {Turner}},\ }\bibfield
		{title} {\bibinfo {title} {Continuous {{Faraday}} measurement of spin
				precession without light shifts},\ }\href
		{https://doi.org/10.1103/PhysRevA.96.063402} {\bibfield  {journal} {\bibinfo
				{journal} {Phys. Rev. A}\ }\textbf {\bibinfo {volume} {96}},\ \bibinfo
			{pages} {063402} (\bibinfo {year} {2017})}\BibitemShut {NoStop}%
		\bibitem [{\citenamefont {Ivakhnenko}\ \emph {et~al.}(2023)\citenamefont
			{Ivakhnenko}, \citenamefont {Shevchenko},\ and\ \citenamefont
			{Nori}}]{ivakhnenko_nonadiabatic_2023}%
		\BibitemOpen
		\bibfield  {author} {\bibinfo {author} {\bibfnamefont {O.~V.}\ \bibnamefont
				{Ivakhnenko}}, \bibinfo {author} {\bibfnamefont {S.~N.}\ \bibnamefont
				{Shevchenko}},\ and\ \bibinfo {author} {\bibfnamefont {F.}~\bibnamefont
				{Nori}},\ }\bibfield  {title} {\bibinfo {title} {Nonadiabatic
				{{Landau}}{\textendash}{{Zener}}{\textendash}{{St{\"u}ckelberg}}{\textendash}{{Majorana}}
				transitions, dynamics, and interference},\ }\href
		{https://doi.org/10.1016/j.physrep.2022.10.002} {\bibfield  {journal}
			{\bibinfo  {journal} {Phys. Rep.}\ }\textbf {\bibinfo {volume} {995}},\
			\bibinfo {pages} {1} (\bibinfo {year} {2023})}\BibitemShut {NoStop}%
		\bibitem [{\citenamefont {Silberfarb}\ \emph {et~al.}(2005)\citenamefont
			{Silberfarb}, \citenamefont {Jessen},\ and\ \citenamefont
			{Deutsch}}]{silberfarb_quantum_2005}%
		\BibitemOpen
		\bibfield  {author} {\bibinfo {author} {\bibfnamefont {A.}~\bibnamefont
				{Silberfarb}}, \bibinfo {author} {\bibfnamefont {P.~S.}\ \bibnamefont
				{Jessen}},\ and\ \bibinfo {author} {\bibfnamefont {I.~H.}\ \bibnamefont
				{Deutsch}},\ }\bibfield  {title} {\bibinfo {title} {Quantum {{State
						Reconstruction}} via {{Continuous Measurement}}},\ }\href
		{https://doi.org/10.1103/PhysRevLett.95.030402} {\bibfield  {journal}
			{\bibinfo  {journal} {Phys. Rev. Lett.}\ }\textbf {\bibinfo {volume} {95}},\
			\bibinfo {pages} {030402} (\bibinfo {year} {2005})}\BibitemShut {NoStop}%
		\bibitem [{\citenamefont {Liu}\ \emph {et~al.}(2009)\citenamefont {Liu},
			\citenamefont {Jung}, \citenamefont {Maxwell}, \citenamefont {Turner},
			\citenamefont {Tiesinga},\ and\ \citenamefont {Lett}}]{liu_quantum_2009-1}%
		\BibitemOpen
		\bibfield  {author} {\bibinfo {author} {\bibfnamefont {Y.}~\bibnamefont
				{Liu}}, \bibinfo {author} {\bibfnamefont {S.}~\bibnamefont {Jung}}, \bibinfo
			{author} {\bibfnamefont {S.~E.}\ \bibnamefont {Maxwell}}, \bibinfo {author}
			{\bibfnamefont {L.~D.}\ \bibnamefont {Turner}}, \bibinfo {author}
			{\bibfnamefont {E.}~\bibnamefont {Tiesinga}},\ and\ \bibinfo {author}
			{\bibfnamefont {P.~D.}\ \bibnamefont {Lett}},\ }\bibfield  {title} {\bibinfo
			{title} {Quantum {{Phase Transitions}} and {{Continuous Observation}} of
				{{Spinor Dynamics}} in an {{Antiferromagnetic Condensate}}},\ }\href
		{https://doi.org/10.1103/PhysRevLett.102.125301} {\bibfield  {journal}
			{\bibinfo  {journal} {Phys. Rev. Lett.}\ }\textbf {\bibinfo {volume} {102}},\
			\bibinfo {pages} {125301} (\bibinfo {year} {2009})}\BibitemShut {NoStop}%
		\bibitem [{Note1()}]{Note1}%
		\BibitemOpen
		\bibinfo {note} {\label {suppmat} See Supplemental Material [url] for
			detailed breakdown of apparatus (I), processing (II), analysis procedure
			(III) and preliminary multi-tone measurements (IV), which includes refs \cite
			{taylor_unambiguous_2023, starkey_scripted_2013, hilbert_grundzuge_1912,
				kornblith_juliadspdspjl_2022, savitzky_smoothing_1964, dierckx_curve_1995,
				bloch_nuclear_1946, ahmad_differential_2022, kelley_4_1999}.}\BibitemShut
		{Stop}%
		\bibitem [{\citenamefont {Vitanov}\ and\ \citenamefont
			{Garraway}(1996)}]{vitanov_landau-zener_1996-1}%
		\BibitemOpen
		\bibfield  {author} {\bibinfo {author} {\bibfnamefont {N.~V.}\ \bibnamefont
				{Vitanov}}\ and\ \bibinfo {author} {\bibfnamefont {B.~M.}\ \bibnamefont
				{Garraway}},\ }\bibfield  {title} {\bibinfo {title} {Landau-{{Zener}} model:
				{{Effects}} of finite coupling duration},\ }\href
		{https://doi.org/10.1103/PhysRevA.53.4288} {\bibfield  {journal} {\bibinfo
				{journal} {Phys. Rev. A}\ }\textbf {\bibinfo {volume} {53}},\ \bibinfo
			{pages} {4288} (\bibinfo {year} {1996})}\BibitemShut {NoStop}%
		\bibitem [{\citenamefont {Zhang}\ and\ \citenamefont
			{Sarovar}(2014)}]{zhang_quantum_2014}%
		\BibitemOpen
		\bibfield  {author} {\bibinfo {author} {\bibfnamefont {J.}~\bibnamefont
				{Zhang}}\ and\ \bibinfo {author} {\bibfnamefont {M.}~\bibnamefont
				{Sarovar}},\ }\bibfield  {title} {\bibinfo {title} {Quantum {{Hamiltonian
						Identification}} from {{Measurement Time Traces}}},\ }\href
		{https://doi.org/10.1103/PhysRevLett.113.080401} {\bibfield  {journal}
			{\bibinfo  {journal} {Phys. Rev. Lett.}\ }\textbf {\bibinfo {volume} {113}},\
			\bibinfo {pages} {080401} (\bibinfo {year} {2014})}\BibitemShut {NoStop}%
		\bibitem [{\citenamefont {{de Clercq}}\ \emph {et~al.}(2016)\citenamefont {{de
					Clercq}}, \citenamefont {Oswald}, \citenamefont {Fl{\"u}hmann}, \citenamefont
			{Keitch}, \citenamefont {Kienzler}, \citenamefont {Lo}, \citenamefont
			{Marinelli}, \citenamefont {Nadlinger}, \citenamefont {Negnevitsky},\ and\
			\citenamefont {Home}}]{de_clercq_estimation_2016}%
		\BibitemOpen
		\bibfield  {author} {\bibinfo {author} {\bibfnamefont {L.~E.}\ \bibnamefont
				{{de Clercq}}}, \bibinfo {author} {\bibfnamefont {R.}~\bibnamefont {Oswald}},
			\bibinfo {author} {\bibfnamefont {C.}~\bibnamefont {Fl{\"u}hmann}}, \bibinfo
			{author} {\bibfnamefont {B.}~\bibnamefont {Keitch}}, \bibinfo {author}
			{\bibfnamefont {D.}~\bibnamefont {Kienzler}}, \bibinfo {author}
			{\bibfnamefont {H.-Y.}\ \bibnamefont {Lo}}, \bibinfo {author} {\bibfnamefont
				{M.}~\bibnamefont {Marinelli}}, \bibinfo {author} {\bibfnamefont
				{D.}~\bibnamefont {Nadlinger}}, \bibinfo {author} {\bibfnamefont
				{V.}~\bibnamefont {Negnevitsky}},\ and\ \bibinfo {author} {\bibfnamefont
				{J.~P.}\ \bibnamefont {Home}},\ }\bibfield  {title} {\bibinfo {title}
			{Estimation of a general time-dependent {{Hamiltonian}} for a single qubit},\
		}\href {https://doi.org/10.1038/ncomms11218} {\bibfield  {journal} {\bibinfo
				{journal} {Nat Commun}\ }\textbf {\bibinfo {volume} {7}},\ \bibinfo {pages}
			{11218} (\bibinfo {year} {2016})}\BibitemShut {NoStop}%
		\bibitem [{\citenamefont {Siva}\ \emph {et~al.}(2023)\citenamefont {Siva},
			\citenamefont {Koolstra}, \citenamefont {Steinmetz}, \citenamefont
			{Livingston}, \citenamefont {Das}, \citenamefont {Chen}, \citenamefont
			{Kreikebaum}, \citenamefont {Stevenson}, \citenamefont {J{\"u}nger},
			\citenamefont {Santiago}, \citenamefont {Siddiqi},\ and\ \citenamefont
			{Jordan}}]{siva_time-dependent_2023}%
		\BibitemOpen
		\bibfield  {author} {\bibinfo {author} {\bibfnamefont {K.}~\bibnamefont
				{Siva}}, \bibinfo {author} {\bibfnamefont {G.}~\bibnamefont {Koolstra}},
			\bibinfo {author} {\bibfnamefont {J.}~\bibnamefont {Steinmetz}}, \bibinfo
			{author} {\bibfnamefont {W.~P.}\ \bibnamefont {Livingston}}, \bibinfo
			{author} {\bibfnamefont {D.}~\bibnamefont {Das}}, \bibinfo {author}
			{\bibfnamefont {L.}~\bibnamefont {Chen}}, \bibinfo {author} {\bibfnamefont
				{J.}~\bibnamefont {Kreikebaum}}, \bibinfo {author} {\bibfnamefont
				{N.}~\bibnamefont {Stevenson}}, \bibinfo {author} {\bibfnamefont
				{C.}~\bibnamefont {J{\"u}nger}}, \bibinfo {author} {\bibfnamefont
				{D.}~\bibnamefont {Santiago}}, \bibinfo {author} {\bibfnamefont
				{I.}~\bibnamefont {Siddiqi}},\ and\ \bibinfo {author} {\bibfnamefont
				{A.}~\bibnamefont {Jordan}},\ }\bibfield  {title} {\bibinfo {title}
			{Time-{{Dependent Hamiltonian Reconstruction Using Continuous Weak
						Measurements}}},\ }\href {https://doi.org/10.1103/PRXQuantum.4.040324}
		{\bibfield  {journal} {\bibinfo  {journal} {PRX Quantum}\ }\textbf {\bibinfo
				{volume} {4}},\ \bibinfo {pages} {040324} (\bibinfo {year}
			{2023})}\BibitemShut {NoStop}%
		\bibitem [{\citenamefont {Naghiloo}\ \emph {et~al.}(2017)\citenamefont
			{Naghiloo}, \citenamefont {Jordan},\ and\ \citenamefont
			{Murch}}]{naghiloo_achieving_2017}%
		\BibitemOpen
		\bibfield  {author} {\bibinfo {author} {\bibfnamefont {M.}~\bibnamefont
				{Naghiloo}}, \bibinfo {author} {\bibfnamefont {A.~N.}\ \bibnamefont
				{Jordan}},\ and\ \bibinfo {author} {\bibfnamefont {K.~W.}\ \bibnamefont
				{Murch}},\ }\bibfield  {title} {\bibinfo {title} {Achieving {{Optimal Quantum
						Acceleration}} of {{Frequency Estimation Using Adaptive Coherent Control}}},\
		}\href {https://doi.org/10.1103/PhysRevLett.119.180801} {\bibfield  {journal}
			{\bibinfo  {journal} {Phys. Rev. Lett.}\ }\textbf {\bibinfo {volume} {119}},\
			\bibinfo {pages} {180801} (\bibinfo {year} {2017})}\BibitemShut {NoStop}%
		\bibitem [{\citenamefont {Tsitouras}(2011)}]{tsitouras_rungekutta_2011}%
		\BibitemOpen
		\bibfield  {author} {\bibinfo {author} {\bibfnamefont {{\relax
						Ch}.}~\bibnamefont {Tsitouras}},\ }\bibfield  {title} {\bibinfo {title}
			{Runge{\textendash}{{Kutta}} pairs of order 5(4) satisfying only the first
				column simplifying assumption},\ }\href
		{https://doi.org/10.1016/j.camwa.2011.06.002} {\bibfield  {journal} {\bibinfo
				{journal} {Comput. Math. with Appl.}\ }\textbf {\bibinfo {volume} {62}},\
			\bibinfo {pages} {770} (\bibinfo {year} {2011})}\BibitemShut {NoStop}%
		\bibitem [{Note2()}]{Note2}%
		\BibitemOpen
		\bibinfo {note} {We use the Julia OrdinaryDifferentialEquations
			ecosystem~\cite {rackauckas_differentialequationsjlperformant_2017} to
			identify the correct basin with a global optimizer~\cite
			{feldt_blackboxoptimjl_2018} before a fast local optimizer~\cite
			{mogensen_optim_2018} locates the minimum and extracts parameter covariances.
			The global optimizer is constrained only by requiring \protect \ensuremath
			{\omega _\protect \text {s}}\protect \, within the span, and positive
			\protect \ensuremath {\Omega _\protect \text {s}}\protect \, less than a
			nominal maximum well into the adiabatic regime. See supplemental material
			\cite {Note1} section III.}\BibitemShut {Stop}%
		\bibitem [{\citenamefont {Majorana}(1932)}]{majorana_atomi_1932}%
		\BibitemOpen
		\bibfield  {author} {\bibinfo {author} {\bibfnamefont {E.}~\bibnamefont
				{Majorana}},\ }\bibfield  {title} {\bibinfo {title} {{Atomi orientati in
					campo magnetico variabile}},\ }\href {https://doi.org/10.1007/BF02960953}
		{\bibfield  {journal} {\bibinfo  {journal} {Nuovo Cim}\ }\textbf {\bibinfo
				{volume} {9}},\ \bibinfo {pages} {43} (\bibinfo {year} {1932})}\BibitemShut
		{NoStop}%
		\bibitem [{\citenamefont {Kofman}\ \emph {et~al.}(2023)\citenamefont {Kofman},
			\citenamefont {Ivakhnenko}, \citenamefont {Shevchenko},\ and\ \citenamefont
			{Nori}}]{kofman_majoranas_2023}%
		\BibitemOpen
		\bibfield  {author} {\bibinfo {author} {\bibfnamefont {P.~O.}\ \bibnamefont
				{Kofman}}, \bibinfo {author} {\bibfnamefont {O.~V.}\ \bibnamefont
				{Ivakhnenko}}, \bibinfo {author} {\bibfnamefont {S.~N.}\ \bibnamefont
				{Shevchenko}},\ and\ \bibinfo {author} {\bibfnamefont {F.}~\bibnamefont
				{Nori}},\ }\bibfield  {title} {\bibinfo {title} {Majorana's approach to
				nonadiabatic transitions validates the adiabatic-impulse approximation},\
		}\href {https://doi.org/10.1038/s41598-023-31084-y} {\bibfield  {journal}
			{\bibinfo  {journal} {Sci Rep}\ }\textbf {\bibinfo {volume} {13}},\ \bibinfo
			{pages} {5053} (\bibinfo {year} {2023})}\BibitemShut {NoStop}%
		\bibitem [{\citenamefont {Zhuang}\ \emph {et~al.}(2022)\citenamefont {Zhuang},
			\citenamefont {Zeng}, \citenamefont {Economou},\ and\ \citenamefont
			{Barnes}}]{zhuang_noise-resistant_2022}%
		\BibitemOpen
		\bibfield  {author} {\bibinfo {author} {\bibfnamefont {F.}~\bibnamefont
				{Zhuang}}, \bibinfo {author} {\bibfnamefont {J.}~\bibnamefont {Zeng}},
			\bibinfo {author} {\bibfnamefont {S.~E.}\ \bibnamefont {Economou}},\ and\
			\bibinfo {author} {\bibfnamefont {E.}~\bibnamefont {Barnes}},\ }\bibfield
		{title} {\bibinfo {title} {Noise-resistant {{Landau-Zener}} sweeps from
				geometrical curves},\ }\href {https://doi.org/10.22331/q-2022-02-02-639}
		{\bibfield  {journal} {\bibinfo  {journal} {Quantum}\ }\textbf {\bibinfo
				{volume} {6}},\ \bibinfo {pages} {639} (\bibinfo {year} {2022})}\BibitemShut
		{NoStop}%
		\bibitem [{Note3()}]{Note3}%
		\BibitemOpen
		\bibinfo {note} {For a weak signal and evolving from an initial eigenstate,
			the Cornu spiral appears eidetically in orthographic projection with one
			focus centered on the pole; kinematically a large Bloch sphere rolls without
			slipping on the planar spiral~\cite {rojo_rolling_2010}.}\BibitemShut {Stop}%
		\bibitem [{\citenamefont {Crameri}\ \emph {et~al.}(2020)\citenamefont
			{Crameri}, \citenamefont {Shephard},\ and\ \citenamefont
			{Heron}}]{crameri_misuse_2020}%
		\BibitemOpen
		\bibfield  {author} {\bibinfo {author} {\bibfnamefont {F.}~\bibnamefont
				{Crameri}}, \bibinfo {author} {\bibfnamefont {G.~E.}\ \bibnamefont
				{Shephard}},\ and\ \bibinfo {author} {\bibfnamefont {P.~J.}\ \bibnamefont
				{Heron}},\ }\bibfield  {title} {\bibinfo {title} {The misuse of colour in
				science communication},\ }\href {https://doi.org/10.1038/s41467-020-19160-7}
		{\bibfield  {journal} {\bibinfo  {journal} {Nat Commun}\ }\textbf {\bibinfo
				{volume} {11}},\ \bibinfo {pages} {5444} (\bibinfo {year}
			{2020})}\BibitemShut {NoStop}%
		\bibitem [{\citenamefont {White}\ \emph {et~al.}()\citenamefont {White},
			\citenamefont {Bounds}, \citenamefont {Taylor}, \citenamefont {Tritt},\ and\
			\citenamefont {Turner}}]{white_unpublished_nodate}%
		\BibitemOpen
		\bibfield  {author} {\bibinfo {author} {\bibfnamefont {S.~J.}\ \bibnamefont
				{White}}, \bibinfo {author} {\bibfnamefont {C.~C.}\ \bibnamefont {Bounds}},
			\bibinfo {author} {\bibfnamefont {H.}~\bibnamefont {Taylor}}, \bibinfo
			{author} {\bibfnamefont {A.}~\bibnamefont {Tritt}},\ and\ \bibinfo {author}
			{\bibfnamefont {L.~D.}\ \bibnamefont {Turner}},\ }\bibfield  {title}
		{\bibinfo {title} {(unpublished)},\ }\href@noop {} {\ }\BibitemShut {NoStop}%
		\bibitem [{\citenamefont {Vitanov}(1999)}]{vitanov_transition_1999}%
		\BibitemOpen
		\bibfield  {author} {\bibinfo {author} {\bibfnamefont {N.~V.}\ \bibnamefont
				{Vitanov}},\ }\bibfield  {title} {\bibinfo {title} {Transition times in the
				{{Landau-Zener}} model},\ }\href {https://doi.org/10.1103/PhysRevA.59.988}
		{\bibfield  {journal} {\bibinfo  {journal} {Phys. Rev. A}\ }\textbf {\bibinfo
				{volume} {59}},\ \bibinfo {pages} {988} (\bibinfo {year} {1999})}\BibitemShut
		{NoStop}%
		\bibitem [{\citenamefont {Colangelo}\ \emph {et~al.}(2017)\citenamefont
			{Colangelo}, \citenamefont {Ciurana}, \citenamefont {Bianchet}, \citenamefont
			{Sewell},\ and\ \citenamefont {Mitchell}}]{colangelo_simultaneous_2017}%
		\BibitemOpen
		\bibfield  {author} {\bibinfo {author} {\bibfnamefont {G.}~\bibnamefont
				{Colangelo}}, \bibinfo {author} {\bibfnamefont {F.~M.}\ \bibnamefont
				{Ciurana}}, \bibinfo {author} {\bibfnamefont {L.~C.}\ \bibnamefont
				{Bianchet}}, \bibinfo {author} {\bibfnamefont {R.~J.}\ \bibnamefont
				{Sewell}},\ and\ \bibinfo {author} {\bibfnamefont {M.~W.}\ \bibnamefont
				{Mitchell}},\ }\bibfield  {title} {\bibinfo {title} {Simultaneous tracking of
				spin angle and amplitude beyond classical limits},\ }\href
		{https://doi.org/10.1038/nature21434} {\bibfield  {journal} {\bibinfo
				{journal} {Nature}\ }\textbf {\bibinfo {volume} {543}},\ \bibinfo {pages}
			{525} (\bibinfo {year} {2017})}\BibitemShut {NoStop}%
		\bibitem [{\citenamefont {Taylor}\ \emph {et~al.}(2023)\citenamefont {Taylor},
			\citenamefont {Bounds}, \citenamefont {Tritt},\ and\ \citenamefont
			{Turner}}]{taylor_unambiguous_2023}%
		\BibitemOpen
		\bibfield  {author} {\bibinfo {author} {\bibfnamefont {H.~A.~M.}\
				\bibnamefont {Taylor}}, \bibinfo {author} {\bibfnamefont {C.~C.}\
				\bibnamefont {Bounds}}, \bibinfo {author} {\bibfnamefont {A.}~\bibnamefont
				{Tritt}},\ and\ \bibinfo {author} {\bibfnamefont {L.~D.}\ \bibnamefont
				{Turner}},\ }\href {https://doi.org/10.48550/arXiv.2309.11825} {\bibinfo
			{title} {Unambiguous measurement in an unshielded microscale magnetometer
				with sensitivity below 1 {{pT}}/{{rHz}}}} (\bibinfo {year} {2023}),\ \Eprint
		{https://arxiv.org/abs/2309.11825} {arxiv:2309.11825 [quant-ph]} \BibitemShut
		{NoStop}%
		\bibitem [{\citenamefont {Starkey}\ \emph {et~al.}(2013)\citenamefont
			{Starkey}, \citenamefont {Billington}, \citenamefont {Johnstone},
			\citenamefont {Jasperse}, \citenamefont {Helmerson}, \citenamefont {Turner},\
			and\ \citenamefont {Anderson}}]{starkey_scripted_2013}%
		\BibitemOpen
		\bibfield  {author} {\bibinfo {author} {\bibfnamefont {P.~T.}\ \bibnamefont
				{Starkey}}, \bibinfo {author} {\bibfnamefont {C.~J.}\ \bibnamefont
				{Billington}}, \bibinfo {author} {\bibfnamefont {S.~P.}\ \bibnamefont
				{Johnstone}}, \bibinfo {author} {\bibfnamefont {M.}~\bibnamefont {Jasperse}},
			\bibinfo {author} {\bibfnamefont {K.}~\bibnamefont {Helmerson}}, \bibinfo
			{author} {\bibfnamefont {L.~D.}\ \bibnamefont {Turner}},\ and\ \bibinfo
			{author} {\bibfnamefont {R.~P.}\ \bibnamefont {Anderson}},\ }\bibfield
		{title} {\bibinfo {title} {A scripted control system for autonomous
				hardware-timed experiments},\ }\href {https://doi.org/10.1063/1.4817213}
		{\bibfield  {journal} {\bibinfo  {journal} {Rev. Sci. Instrum.}\ }\textbf
			{\bibinfo {volume} {84}},\ \bibinfo {pages} {085111} (\bibinfo {year}
			{2013})}\BibitemShut {NoStop}%
		\bibitem [{\citenamefont {Hilbert}(1912)}]{hilbert_grundzuge_1912}%
		\BibitemOpen
		\bibfield  {author} {\bibinfo {author} {\bibfnamefont {D.}~\bibnamefont
				{Hilbert}},\ }\href@noop {} {\emph {\bibinfo {title} {{Grundz{\"u}ge einer
						allgemeinen theorie der linearen integralgleichungen}}}}\ (\bibinfo
		{publisher} {{Leipzig, B. G. Teubner}},\ \bibinfo {year} {1912})\BibitemShut
		{NoStop}%
		\bibitem [{\citenamefont {Kornblith}\ \emph {et~al.}(2022)\citenamefont
			{Kornblith}, \citenamefont {Lynch}, \citenamefont {Holters}, \citenamefont
			{Santos}, \citenamefont {Russell}, \citenamefont {Kickliter}, \citenamefont
			{Bezanson}, \citenamefont {Adalsteinsson}, \citenamefont {Arslan},
			\citenamefont {Yamamoto}, \citenamefont {{jordancluts}}, \citenamefont
			{Shah}, \citenamefont {Pastell}, \citenamefont {Kelman}, \citenamefont
			{Arthur}, \citenamefont {Krauss}, \citenamefont {HDictus}, \citenamefont
			{{El-Saawy}}, \citenamefont {Kofron}, \citenamefont {Hanson}, \citenamefont
			{Luke}, \citenamefont {Kastanos}, \citenamefont {Bill}, \citenamefont
			{Clemens}, \citenamefont {Saba}, \citenamefont {{ibadr}}, \citenamefont
			{Bolewski},\ and\ \citenamefont {Smith}}]{kornblith_juliadspdspjl_2022}%
		\BibitemOpen
		\bibfield  {author} {\bibinfo {author} {\bibfnamefont {S.}~\bibnamefont
				{Kornblith}}, \bibinfo {author} {\bibfnamefont {G.}~\bibnamefont {Lynch}},
			\bibinfo {author} {\bibfnamefont {M.}~\bibnamefont {Holters}}, \bibinfo
			{author} {\bibfnamefont {J.~F.}\ \bibnamefont {Santos}}, \bibinfo {author}
			{\bibfnamefont {S.}~\bibnamefont {Russell}}, \bibinfo {author} {\bibfnamefont
				{J.}~\bibnamefont {Kickliter}}, \bibinfo {author} {\bibfnamefont
				{J.}~\bibnamefont {Bezanson}}, \bibinfo {author} {\bibfnamefont
				{G.}~\bibnamefont {Adalsteinsson}}, \bibinfo {author} {\bibfnamefont
				{A.}~\bibnamefont {Arslan}}, \bibinfo {author} {\bibfnamefont
				{R.}~\bibnamefont {Yamamoto}}, \bibinfo {author} {\bibnamefont
				{{jordancluts}}}, \bibinfo {author} {\bibfnamefont {V.~B.}\ \bibnamefont
				{Shah}}, \bibinfo {author} {\bibfnamefont {M.}~\bibnamefont {Pastell}},
			\bibinfo {author} {\bibfnamefont {T.}~\bibnamefont {Kelman}}, \bibinfo
			{author} {\bibfnamefont {B.}~\bibnamefont {Arthur}}, \bibinfo {author}
			{\bibfnamefont {T.}~\bibnamefont {Krauss}}, \bibinfo {author} {\bibnamefont
				{HDictus}}, \bibinfo {author} {\bibfnamefont {H.}~\bibnamefont {{El-Saawy}}},
			\bibinfo {author} {\bibfnamefont {J.}~\bibnamefont {Kofron}}, \bibinfo
			{author} {\bibfnamefont {E.}~\bibnamefont {Hanson}}, \bibinfo {author}
			{\bibfnamefont {R.}~\bibnamefont {Luke}}, \bibinfo {author} {\bibfnamefont
				{A.}~\bibnamefont {Kastanos}}, \bibinfo {author} {\bibnamefont {Bill}},
			\bibinfo {author} {\bibnamefont {Clemens}}, \bibinfo {author} {\bibfnamefont
				{E.}~\bibnamefont {Saba}}, \bibinfo {author} {\bibnamefont {{ibadr}}},
			\bibinfo {author} {\bibfnamefont {J.}~\bibnamefont {Bolewski}},\ and\
			\bibinfo {author} {\bibfnamefont {J.}~\bibnamefont {Smith}},\ }\href
		{https://doi.org/10.5281/zenodo.7406426} {\bibinfo {title}
			{{{JuliaDSP}}/{{DSP}}.jl: V0.7.8}},\ \bibinfo {howpublished} {Zenodo}
		(\bibinfo {year} {2022})\BibitemShut {NoStop}%
		\bibitem [{\citenamefont {Savitzky}\ and\ \citenamefont
			{Golay}(1964)}]{savitzky_smoothing_1964}%
		\BibitemOpen
		\bibfield  {author} {\bibinfo {author} {\bibfnamefont {{\relax
						Abraham}.}~\bibnamefont {Savitzky}}\ and\ \bibinfo {author} {\bibfnamefont
				{M.~J.~E.}\ \bibnamefont {Golay}},\ }\bibfield  {title} {\bibinfo {title}
			{Smoothing and {{Differentiation}} of {{Data}} by {{Simplified Least Squares
						Procedures}}.},\ }\href {https://doi.org/10.1021/ac60214a047} {\bibfield
			{journal} {\bibinfo  {journal} {Anal. Chem.}\ }\textbf {\bibinfo {volume}
				{36}},\ \bibinfo {pages} {1627} (\bibinfo {year} {1964})}\BibitemShut
		{NoStop}%
		\bibitem [{\citenamefont {Dierckx}(1995)}]{dierckx_curve_1995}%
		\BibitemOpen
		\bibfield  {author} {\bibinfo {author} {\bibfnamefont {P.}~\bibnamefont
				{Dierckx}},\ }\href@noop {} {\emph {\bibinfo {title} {Curve and {{Surface
							Fitting}} with {{Splines}}}}}\ (\bibinfo  {publisher} {{Clarendon Press}},\
		\bibinfo {year} {1995})\BibitemShut {NoStop}%
		\bibitem [{\citenamefont {Bloch}(1946)}]{bloch_nuclear_1946}%
		\BibitemOpen
		\bibfield  {author} {\bibinfo {author} {\bibfnamefont {F.}~\bibnamefont
				{Bloch}},\ }\bibfield  {title} {\bibinfo {title} {Nuclear {{Induction}}},\
		}\href {https://doi.org/10.1103/PhysRev.70.460} {\bibfield  {journal}
			{\bibinfo  {journal} {Phys. Rev.}\ }\textbf {\bibinfo {volume} {70}},\
			\bibinfo {pages} {460} (\bibinfo {year} {1946})}\BibitemShut {NoStop}%
		\bibitem [{\citenamefont {Ahmad}\ \emph {et~al.}(2022)\citenamefont {Ahmad},
			\citenamefont {Isa}, \citenamefont {Lim},\ and\ \citenamefont
			{Ang}}]{ahmad_differential_2022}%
		\BibitemOpen
		\bibfield  {author} {\bibinfo {author} {\bibfnamefont {M.~F.}\ \bibnamefont
				{Ahmad}}, \bibinfo {author} {\bibfnamefont {N.~A.~M.}\ \bibnamefont {Isa}},
			\bibinfo {author} {\bibfnamefont {W.~H.}\ \bibnamefont {Lim}},\ and\ \bibinfo
			{author} {\bibfnamefont {K.~M.}\ \bibnamefont {Ang}},\ }\bibfield  {title}
		{\bibinfo {title} {Differential evolution: {{A}} recent review based on
				state-of-the-art works},\ }\href {https://doi.org/10.1016/j.aej.2021.09.013}
		{\bibfield  {journal} {\bibinfo  {journal} {Alex. Eng. J.}\ }\textbf
			{\bibinfo {volume} {61}},\ \bibinfo {pages} {3831} (\bibinfo {year}
			{2022})}\BibitemShut {NoStop}%
		\bibitem [{\citenamefont {Kelley}(1999)}]{kelley_4_1999}%
		\BibitemOpen
		\bibfield  {author} {\bibinfo {author} {\bibfnamefont {C.~T.}\ \bibnamefont
				{Kelley}},\ }\bibfield  {title} {\bibinfo {title} {4. {{The BFGS Method}}},\
		}in\ \href {https://doi.org/10.1137/1.9781611970920.ch4} {\emph {\bibinfo
				{booktitle} {Iterative {{Methods}} for {{Optimization}}}}},\ \bibinfo {series
			and number} {Front. {{Appl}}. {{Math}}. {{Stat}}.}\ (\bibinfo  {publisher}
		{{SIAM}},\ \bibinfo {year} {1999})\ pp.\ \bibinfo {pages}
		{71--86}\BibitemShut {NoStop}%
		\bibitem [{\citenamefont {Rackauckas}\ and\ \citenamefont
			{Nie}(2017)}]{rackauckas_differentialequationsjlperformant_2017}%
		\BibitemOpen
		\bibfield  {author} {\bibinfo {author} {\bibfnamefont {C.}~\bibnamefont
				{Rackauckas}}\ and\ \bibinfo {author} {\bibfnamefont {Q.}~\bibnamefont
				{Nie}},\ }\bibfield  {title} {\bibinfo {title}
			{{{DifferentialEquations}}.jl{\textendash}a performant and feature-rich
				ecosystem for solving differential equations in {{Julia}}},\ }\href
		{https://doi.org/10.5334/jors.151} {\bibfield  {journal} {\bibinfo  {journal}
				{J. Open Res. Softw.}\ }\textbf {\bibinfo {volume} {5}},\ \bibinfo {pages}
			{15} (\bibinfo {year} {2017})}\BibitemShut {NoStop}%
		\bibitem [{\citenamefont {Feldt}(2018)}]{feldt_blackboxoptimjl_2018}%
		\BibitemOpen
		\bibfield  {author} {\bibinfo {author} {\bibfnamefont {R.}~\bibnamefont
				{Feldt}},\ }\href@noop {} {\bibinfo {title} {{{BlackBoxOptim}}.jl}},\
		\bibinfo {howpublished} {GitHub Repository} (\bibinfo {year}
		{2018})\BibitemShut {NoStop}%
		\bibitem [{\citenamefont {Mogensen}\ and\ \citenamefont
			{Riseth}(2018)}]{mogensen_optim_2018}%
		\BibitemOpen
		\bibfield  {author} {\bibinfo {author} {\bibfnamefont {P.~K.}\ \bibnamefont
				{Mogensen}}\ and\ \bibinfo {author} {\bibfnamefont {A.~N.}\ \bibnamefont
				{Riseth}},\ }\bibfield  {title} {\bibinfo {title} {Optim: {{A}} mathematical
				optimization package for {{Julia}}},\ }\href
		{https://doi.org/10.21105/joss.00615} {\bibfield  {journal} {\bibinfo
				{journal} {J. Open Source Softw.}\ }\textbf {\bibinfo {volume} {3}},\
			\bibinfo {pages} {615} (\bibinfo {year} {2018})}\BibitemShut {NoStop}%
		\bibitem [{\citenamefont {Rojo}\ and\ \citenamefont
			{Bloch}(2010)}]{rojo_rolling_2010}%
		\BibitemOpen
		\bibfield  {author} {\bibinfo {author} {\bibfnamefont {A.~G.}\ \bibnamefont
				{Rojo}}\ and\ \bibinfo {author} {\bibfnamefont {A.~M.}\ \bibnamefont
				{Bloch}},\ }\bibfield  {title} {\bibinfo {title} {The rolling sphere, the
				quantum spin, and a simple view of the {{Landau}}{\textendash}{{Zener}}
				problem},\ }\href {https://doi.org/10.1119/1.3456565} {\bibfield  {journal}
			{\bibinfo  {journal} {Am. J. Phys.}\ }\textbf {\bibinfo {volume} {78}},\
			\bibinfo {pages} {1014} (\bibinfo {year} {2010})}\BibitemShut {NoStop}%
	\end{thebibliography}
\end{document}